\documentclass[%
 reprint,
 amsmath,amssymb,
 aps,
superscriptaddress, 
]{revtex4-2}
\usepackage{siunitx}
\usepackage{appendix}
\usepackage{graphicx}% Include figure files
\usepackage{dcolumn}% Align table columns on decimal point
\usepackage{bm}% bold math
\usepackage{hyperref}
\usepackage{xcolor}
\usepackage[linesnumbered,ruled,vlined]{algorithm2e}
\usepackage{algorithmic}
\usepackage{listings}
\usepackage{amsmath}

\renewcommand{\eqref}[1]{Eq. (\textup{\ref{#1}})}

\SetKwInput{KwInput}{Input}                % Set the Input

\begin{document}

\title{Online learning of the transfer matrix of dynamic scattering media: wavefront shaping meets multidimensional time series}
\author{Lorenzo Valzania}
\thanks{Correspondence to: lorenzo.valzania1@gmail.com,\\sylvain.gigan@lkb.ens.fr}
\affiliation{Laboratoire Kastler Brossel, \'Ecole Normale Sup\'erieure - Paris Sciences et Lettres (PSL) Research University, Sorbonne Universit\'e, Centre National de la Recherche Scientifique (CNRS) UMR 8552, Coll\`ege de France, 24 rue Lhomond, 75005 Paris, France}
\author{Sylvain Gigan}
\thanks{Correspondence to: lorenzo.valzania1@gmail.com,\\sylvain.gigan@lkb.ens.fr}
\affiliation{Laboratoire Kastler Brossel, \'Ecole Normale Sup\'erieure - Paris Sciences et Lettres (PSL) Research University, Sorbonne Universit\'e, Centre National de la Recherche Scientifique (CNRS) UMR 8552, Coll\`ege de France, 24 rue Lhomond, 75005 Paris, France}

\date{\today}
% It is always \today, today,
             %  but any date may be explicitly specified
%% To be edited by editor
% \dates{Compiled \today}

%\ociscodes{(140.3490) Lasers, distributed feedback; (060.2420) Fibers, polarization-maintaining;(060.3735) Fiber Bragg gratings.}

%% To be edited by editor
% \doi{\url{http://dx.doi.org/10.1364/XX.XX.XXXXXX}}

\begin{abstract}
\noindent
% Long and multidimensional time series hold a central place in the study of biological systems.
Thanks to the latest advancements in wavefront shaping, optical methods have proven crucial to achieve imaging and control light in multiply scattering media, like biological tissues. However, the stability times of living biological specimens often prevent such methods from gaining insights into relevant functioning mechanisms in cellular and organ systems. Here we present a recursive and online optimization routine, borrowed from time series analysis, to optimally track the transfer matrix of dynamic scattering media over arbitrarily long timescales. While preserving the advantages of both optimization-based routines and transfer-matrix measurements, it operates in a memory-efficient manner. Because it can be readily implemented in existing wavefront shaping setups, featuring amplitude and/or phase modulation and phase-resolved or intensity-only acquisition, it paves the way for efficient optical investigations of living biological specimens.
\end{abstract}

\maketitle

%%%%%%%%%%%%%%%%%%%%%%%%%%%%%%%%%%%%%%%%%%%%%%%%%%%%
% Although there are no explicit sections in letters, we still divide the content in sections for a better organization. They will be removed at the end.
%%%%%%%%%%%%%%%%%%%%%%%%%%%%%%%%%%%%%%%%%%%%%%%%%%%%

%%%%%%%%%%%%%%%%%%%%%%%%%%%%%%%%%%%%%%%%%%%%%%%%%%%%
%%%%%%%%%%%%%%%%%%  INTRODUCTION  %%%%%%%%%%%%%%%%%%
%%%%%%%%%%%%%%%%%%%%%%%%%%%%%%%%%%%%%%%%%%%%%%%%%%%%

\section{Introduction}
\label{sec:intro}
\noindent
% Understanding the functioning mechanisms of cells and organisms, in order to predict their behaviour, requires their observation at extremely different timescales, from nanoseconds (at a molecular level) to minutes (for organ systems) \cite{assmus2006dynamics}. This often relies on solving inverse problems from long and multidimensional time series \cite{engl2009inverse}, whose prohibitive size can make their evaluation problematic. Moreover, the need for fast data acquisitions results, in turn, in measurements with inherently low signal-to-noise ratios.

Optical methods are an irreplaceable tool to investigate biological media. They deliver images at numerous contrast mechanisms \cite{mertz2019introduction}, and can activate injected biomolecules \cite{ellis2007caged} and fluorescent markers \cite{lichtman2005fluorescence}. However, precisely delivering light in space and time through biological tissues is not straightforward, as photons get multiply scattered by heterogeneities of tissues, limiting their penetration depth \cite{rotter2017light}.

Another current challenge lies in tracking the scattering behaviour of living specimens, with decorrelation times up to only a few ms \cite{liu2015optical}. This proves crucial to understand the functioning mechanisms of cells and organisms, which requires their observation at extremely different timescales, from nanoseconds (at a molecular level) to minutes (for organ systems) \cite{assmus2006dynamics}.
The need for fast data acquisitions results, in turn, in measurements with inherently low signal-to-noise ratios, and requires solving long and multidimensional time series \cite{engl2009inverse}, whose prohibitive size can make their evaluation problematic.

Wavefront shaping techniques have established themselves as the tools of choice to guide light in scattering media \cite{vellekoop2015feedback}. The transmission of arbitrary fields \cite{devaud2021speckle}, point-spread-function (PSF) engineering \cite{boniface2017transmission}, imaging \cite{popoff2010image}, as well as tuning energy transmission through scattering media \cite{kim2012maximal}, become all accessible if the transfer matrix of the medium is measured \cite{popoff2010measuring, vellekoop2015feedback}. However, conventional methods to retrieve the transfer matrix yield sub-optimal solutions in noisy environments \cite{vellekoop2015feedback}. Those optimization routines which can compensate for noise in the transfer matrix \cite{matthes2021learning}, however, require storing in memory the whole history of past measurements, making them unsuited with long streams of data.

Iterative, optimization-based, sequential algorithms to focus through scattering media yield an increase in the focus intensity already at their early iterations, which makes them the preferred option on dynamic media. Importantly, they are cast as \textit{recursive} procedures, \textit{i.e.}, computing the new estimate of the solution only requires the previous estimate and the new data point. Unfortunately, their stochastic nature makes optimization over a set of output modes less reliable and the transmission of arbitrary fields prohibitive. Moreover, these procedures rely on maximizing a given metric, limiting light control to one predefined task. Various implementations derived from genetic algorithms \cite{conkey2012genetic, cheng2022long} have shown better resilience to noise than sequential algorithms, however at the cost of a higher computational complexity and careful choice of several adjustable parameters.
% Furthermore, a framework to deal with the characterization of dynamic scattering media is currently lacking.

In signal processing, communications and finance, where most datasets are multidimensional time series, the recursive least-squares (RLS) algorithm has played a central role for system identification and prediction \cite{haykin2014adaptive, haykin1997adaptive, patra2017adaptive}. It allows optimal learning of linear predictors in an online manner---predictors are updated every time a new piece of data is sequentially made available, however past data do not need to be stored in memory. Consequently, its computational complexity is independent of the length of the time series, so iterations can be run over and over, ideally at the same rate as data acquisition (real-time operation).

Here, we demonstrate that the RLS algorithm represents a valuable tool to optimally estimate the transfer matrix of dynamic scattering media online and recursively. The least-squares optimization ensures resilience to noise. The algorithm is provided with a tunable memory, such that the dynamics of the scattering medium is accounted for. By doing so only the most reliable data points, \textit{i.e.}, those acquired within the stability time of the medium, are used during the optimization. We justify how the RLS model can fit a wide variety of dynamic mechanisms happening in scattering media. Its performance is showcased with both simulated and experimental results, tracking the transmission matrix and the time-gated reflection matrix at realistic noise levels and stability times. We further show how light optimization can be achieved with binary amplitude or phase modulation and with phase-resolved or intensity-only measurements. Based on its computational complexity, we discuss its feasibility for light control in living biological specimens at large fields of view. Its simple implementation and the low number of adjustable parameters (whose choice is motivated in the next sections) make our proposed method readily applicable in existing wavefront shaping setups.

%%%%%%%%%%%%%%%%%%%%%%%%%%%%%%%%%%%%%%%%%%%%%%%%%%%%
%%%%%%%%%%%%%%%%%%  METHODS  %%%%%%%%%%%%%%%%%%
%%%%%%%%%%%%%%%%%%%%%%%%%%%%%%%%%%%%%%%%%%%%%%%%%%%

\section{Methods}
\label{sec:methods}
\begin{figure*}[htb]
    \centering\includegraphics[width=\textwidth]{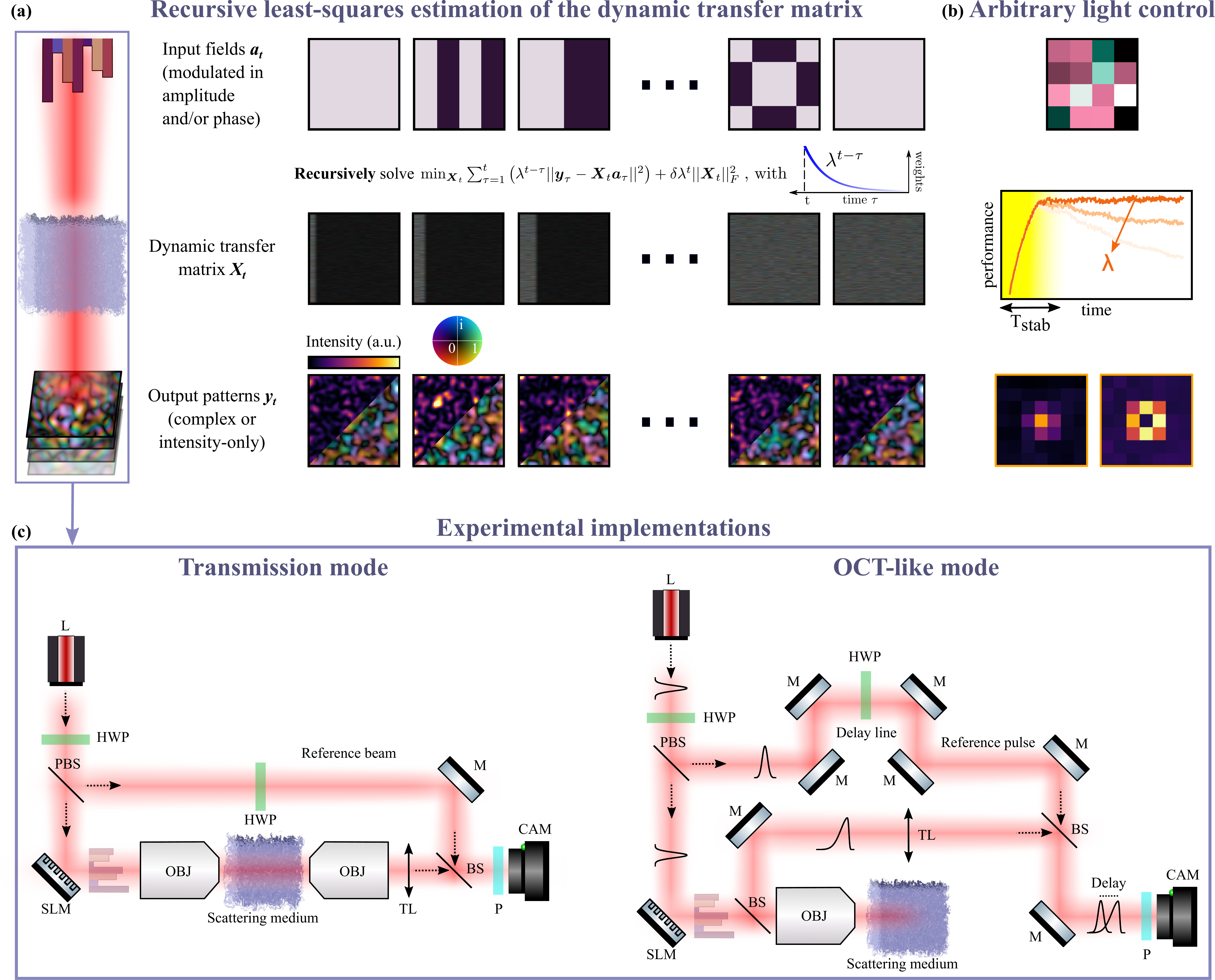}
    \caption{Graphical summary of the RLS estimation technique and experimental implementations. (a) A sequence of input fields, modulated in amplitude and/or in phase (here Hadamard modulation patterns are shown), interacts with a dynamic scattering medium with unknown transfer matrix. Each (input, output) pair is used to update recursively the estimation of the dynamic transfer matrix, minimizing a linear least-squares loss function, where each term is weighted via the coefficient $\lambda^{t-\tau}$. (b) At every time step, upon optimizing the coefficient $\lambda$, the current estimate of the transfer matrix can be used to achieve arbitrary light control through the scattering medium (here, a focus and a donut-shaped beam are displayed). (c) We demonstrated our method with a setup in transmission mode, for the retrieval of the transmission matrix (left), and with an OCT setup, for the retrieval of the time-gated reflection matrix (right). L: laser source; HWP: half-wave plate; (P)BS: (polarizing) beam-splitter; SLM: liquid-crystal-based spatial light modulator; OBJ: objective lens; TL: tube lens; P: polarizer; M: mirror; CAM: camera.}
    \label{fig_method_summary}
\end{figure*}

\noindent
The method bears similarities with conventional routines for the measurement of the transfer matrix, and its working principle is graphically summarized in Fig. \ref{fig_method_summary}(a).
However, here we allow the transfer matrix $\boldsymbol{X}_t \in \mathbb{C}^{M \times N}$ of the scattering medium to be dynamic, where we have denoted the number of output and input degrees of freedom with $M$ and $N$, respectively. At every time step $t$, while probing the medium with the input $\boldsymbol{a}_t \in \mathbb{C}^{N}$ and collecting the corresponding output $\boldsymbol{y}_t = \boldsymbol{X}_t \boldsymbol{a}_t \in \mathbb{C}^{M}$, we aim to solve the optimization problem $\boldsymbol{\hat{X}}_t = \arg \min_{\boldsymbol{X}_t} \mathcal{L}_t(\boldsymbol{X}_t)$, with
\begin{equation}
    \mathcal{L}_t(\boldsymbol{X}_t) \equiv \sum_{\tau = 1}^{t} \left( \lambda^{t-\tau} ||\boldsymbol{y}_\tau - \boldsymbol{X}_t \boldsymbol{a}_\tau ||^2 \right) + \delta \lambda^t ||\boldsymbol{X}_t||_F^2 \, ,
    \label{eq:loss}
\end{equation}
and where $||\cdot||$ and $||\cdot||_F$ denote the $L^2$-norm of a vector and the Frobenius norm of a matrix, respectively. Although for sake of generality the inputs and the outputs are assumed to be complex, we will also report an implementation where they are real, meaning that only the amplitude of the input beam is modulated and the intensity of the output fields is measured. Equation \ref{eq:loss} is a linear least-squares loss function, featuring Tikhonov regularization via the regularization constant $\delta$. Note, however, that each \textit{data-fidelity term} $||\boldsymbol{y}_\tau - \boldsymbol{X}_t \boldsymbol{a}_\tau ||^2$ is exponentially weighted in time, such that the old pieces of data (corresponding to $\tau \ll t$) are less relevant than the most recent ones in the current estimation of the transfer matrix at time $t$. In other words, the \textit{forgetting factor} $\lambda \leq 1$ endows the algorithm with a memory, which allows it to cope with dynamic transfer matrices---at every time step $t$, the optimization problem is solved anew, using the whole history of past data, where more contribution is given to newest data. Evidently, in the case of a static scattering medium, all measurements can be equally trusted, thus \eqref{eq:loss} reduces to a typical regularized linear least-squares problem upon setting $\lambda = 1$. Once $\lambda$ and $\delta$ are fixed, the least-squares problem has a unique solution, provided the inputs are linearly independent, which is the case in conventional transfer-matrix measurements, where the inputs are drawn from the Hadamard basis of order $N$.

The choice of exponential weights for \eqref{eq:loss} is motivated by the physics of our problem. We aim to follow the evolution of the transfer matrix of dynamic scattering media, subjected to uncorrelated variations, whereby the total transferred power fraction is constant in time. These conditions apply in a wide variety of dynamic mechanisms in scattering media investigated with visible and near-infrared light, \textit{e.g.} whenever their inner scatterers move due to functional changes \cite{brake2016analyzing, liu2015optical}, or even when the sample drifts away from its initial position, suggesting that our method can also be used as an online calibration tool of imaging systems. In all these situations, the transfer matrix can indeed be described by the time series \cite{vellekoop2008phase},
\begin{equation}
    \boldsymbol{X}_{t} = \frac{\sigma_{\boldsymbol{X}}}{\sqrt{\sigma_{\boldsymbol{X}}^2 + \sigma_{\boldsymbol{P}}^2}}(\boldsymbol{X}_{t-1} + \boldsymbol{P}_{t}) \, ,
    \label{eq:ar1}
\end{equation}
where we assume that both the transfer matrix and the perturbation matrix $\boldsymbol{P}_t$ are random variables independently drawn from complex Gaussian distributions with zero mean and constant variance $\sigma_{\boldsymbol{X}}^2$ and $\sigma_{\boldsymbol{P}}^2$, respectively \cite{webster2004spectral}. Equation \ref{eq:ar1} denotes an autoregressive model of order 1, AR(1), whose autocovariance is proportional to $(\sigma_{\boldsymbol{X}} / \sqrt{\sigma_{\boldsymbol{X}}^2 + \sigma_{\boldsymbol{P}}^2})^t$, justifying our exponentially weighted model of \eqref{eq:loss}. When focusing through dynamic scattering media following \eqref{eq:ar1}, the stability time of the enhancement is proportional to $\sigma_{\boldsymbol{P}}^{-2}$ \cite{vellekoop2008phase}. This means that the optimal weight $\lambda$ should follow the same dependence, thus in principle requiring the knowledge of the rate of change of the scattering medium. A strategy for automatically tuning the forgetting factor will be discussed in section \ref{sec:results}.

Crucially, minimizing the loss function of \eqref{eq:loss} does not require storing the whole history of past data. This becomes apparent if we recall that the linear least-squares estimate of $\boldsymbol{X}_t$, $\boldsymbol{\hat{X}}_t$, satisfies the normal equations,
\begin{equation}
    \boldsymbol{C}_t \boldsymbol{\hat{X}}_t^H = \boldsymbol{K}_t \, ,
    \label{eq:normal}
\end{equation}
with the covariance matrix of inputs and the cross-covariance matrix at time $t$ respectively defined as,
\begin{subequations}
    \begin{equation}
        \boldsymbol{C}_t \equiv \sum_{\tau = 1}^{t} \left( \lambda^{t-\tau} \boldsymbol{a}_\tau \boldsymbol{a}_\tau^H \right) + \delta \lambda^t \boldsymbol{I}_N \in \mathbb{C}^{N \times N}
        \label{eq:covariance_def}
    \end{equation}
    \begin{equation}
        \boldsymbol{K}_t \equiv \sum_{\tau = 1}^{t} \lambda^{t-\tau} \boldsymbol{a}_\tau \boldsymbol{y}_\tau^H \in \mathbb{C}^{N \times M} \, ,
        \label{eq:cross_covariance_def}
    \end{equation}
    \label{eq:tow_covariances_def}
\end{subequations}
with $\boldsymbol{I}_N$ denoting the identity matrix of order $N$ and the superscript $H$ standing for Hermitian transposition. The quantities calculated in Eqs. (\ref{eq:tow_covariances_def}) can be both estimated recursively, as follows:
\begin{subequations}
    \begin{equation}
        \boldsymbol{C}_t = \lambda \boldsymbol{C}_{t-1} + \boldsymbol{a}_t \boldsymbol{a}_t^H
    \label{eq:covariance_recursive}
    \end{equation}
    \begin{equation}
        \boldsymbol{K}_t = \lambda \boldsymbol{K}_{t-1} + \boldsymbol{a}_t \boldsymbol{y}_t^H \, .
    \label{eq:cross_covariance_recursive}
    \end{equation}
    \label{eq:two_covariances_recursive}
\end{subequations}
Equations (\ref{eq:two_covariances_recursive}) mean the right-hand side of \eqref{eq:loss} can be minimized from the previous estimates of the covariance and cross-covariance matrices and the new piece of data $(\boldsymbol{a}_t, \boldsymbol{y}_t)$. It becomes now clear how the RLS algorithm combines the benefits of transfer-matrix-based and optimization approaches. Using a recursive procedure, a typical asset of, \textit{e.g.}, the continuous sequential algorithm (CSA), the partitioning algorithm \cite{vellekoop2008phase}, or more computationally intense genetic algorithms \cite{conkey2012genetic}, the full $\boldsymbol{X}_t$ is estimated \textit{in parallel} at all output pixels, thereby preserving all light-control capabilities allowed by the knowledge of the transfer matrix \cite{popoff2010image, popoff2010image, matthes2021learning, kim2012maximal, boniface2017transmission} [Fig. \ref{fig_method_summary}(b)]. In principle, the transfer matrix could be obtained from \eqref{eq:normal} as $\boldsymbol{\hat{X}}_t = \boldsymbol{K}_t^H (\boldsymbol{C}_{t}^{-1})^H$. However, in what follows we will implement the \textit{inverse QR-decomposition-based RLS} (abbreviated as inverse QRD-RLS) algorithm \cite{alexander1993method}. Because it avoids matrix inversions and it always preserves the non-negativeness of the covariance matrix, it possesses higher numerical stability than directly inverting \eqref{eq:normal}. Overall, it boils down to performing a QR decomposition of a matrix constructed from the new data and the previous estimate of the square root of the inverse covariance matrix. This results in few lines of code which can be readily implemented in any programming language using standard libraries or built-in functions (see the box Algorithm \ref{algo:RLS} and the corresponding code available at Ref. \cite{github_code}). As can be seen from \eqref{eq:loss} and Algorithm \ref{algo:RLS}, the regularization constant $\delta$ is used to construct the initial estimate of the square root of the inverse correlation matrix, hence it mostly impacts the convergence speed at early iterations. In section \ref{sec:results}, the choice of its value will be discussed.

%%%%%%%%%%%%%%%%%%%%%%%%%%%%%%%%%%%%%%%%%%%%%%%%%%%%
%%%%%%%%%%%%%%%%%%  ALGORITHM  %%%%%%%%%%%%%%%%%%
%%%%%%%%%%%%%%%%%%%%%%%%%%%%%%%%%%%%%%%%%%%%%%%%%%%%
%
\begin{algorithm}[htbp]
    \DontPrintSemicolon
    \caption{Inverse QRD-RLS update\\\small{Initializations: $\boldsymbol{\hat{X}}_{0} = \boldsymbol{0}$, $(\boldsymbol{C}_{0}^{-1})^{1/2} = \delta^{-1/2} \boldsymbol{I}_N$}}
    \label{algo:RLS}
    \KwInput{New input pattern $\boldsymbol{a}_t$, new output pattern $\boldsymbol{y}_t$, previous estimate of the transfer matrix $\boldsymbol{\hat{X}}_{t-1}$, previous estimate of the square root of the inverse covariance matrix $(\boldsymbol{C}_{t-1}^{-1})^{1/2}$, forgetting factor $\lambda$}
    \tcc{Construction of the matrix $\boldsymbol{U}$}
    $\boldsymbol{U} =
    \begin{bmatrix}
        1 & \lambda^{-1/2} \boldsymbol{a}_t^H (\boldsymbol{C}_{t-1}^{-1})^{1/2}\\
        \boldsymbol{0} &  \lambda^{-1/2} (\boldsymbol{C}_{t-1}^{-1})^{1/2} 
    \end{bmatrix}$\\
    \tcc{QR decomposition of $\boldsymbol{U}^H$}
    $\boldsymbol{U}^H = \boldsymbol{Q} \boldsymbol{V}^H$\\
    $\boldsymbol{V} = 
    \begin{bmatrix}
        v_{11} & \boldsymbol{0}^H\\
        \boldsymbol{v}_{21} & (\boldsymbol{C}_{t}^{-1})^{1/2}
    \end{bmatrix}$\\
    \tcc{Update of the transfer matrix}
    $\boldsymbol{\hat{X}}_t = \boldsymbol{\hat{X}}_{t-1} + (\boldsymbol{y}_t - \boldsymbol{\hat{X}}_{t-1} \boldsymbol{a}_t) \boldsymbol{v}_{21}^H v_{11}^{-1}$\\
    \textbf{return} $\boldsymbol{\hat{X}}_t$ and $(\boldsymbol{C}_{t}^{-1})^{1/2}$
\end{algorithm}
%
%%%%%%%%%%%%%%%%%%%%%%%%%%%%%%%%%%%%%%%%%%%%%%%%%%%%
%%%%%%%%%%%%%%%%%%  EXPERIMENTS  %%%%%%%%%%%%%%%%%%
%%%%%%%%%%%%%%%%%%%%%%%%%%%%%%%%%%%%%%%%%%%%%%%%%%%%
\section{Experiments}
\label{sec:experiments}
\noindent
Figure \ref{fig_method_summary}(c) shows the sketches of the experimental implementations used to demonstrate our method. Both are based on phase-shifting digital holography to retrieve the complex output fields $\boldsymbol{y}_t$ after interacting with a multiply scattering medium. The medium is an opaque deposit of ZnO nanoparticles (size $<$ 100 nm), whose thickness (20 $\si{\micro\meter}$) is $\sim$5 transport mean free paths, ensuring full mixing of its optical modes at the output. The input fields are shaped via a reflective, phase-only and liquid-crystal-based spatial light modulator (SLM, Meadowlark Optics HSP512L-1064) and focused on the scattering medium with an objective with a numerical aperture of 0.4 (Olympus PLN20X). A region-of-interest containing $\sim$80 speckle grains is imaged onto a CCD camera (Manta G-046B, Allied Vision) via a tube lens, yielding a pixel size of 0.2 $\si{\micro\meter}$ at the CCD plane. Before impinging onto the SLM, part of the beam is redirected along a reference arm with a polarizing beam splitter (PBS), and subsequently recombined with the scattered beam through a beam splitter (BS). The relative power of the two beams, yielding the maximum interference contrast, is adjusted via two half-wave plates, one along the common path and one along the reference arm, while a polarizer in front of the camera filters out any potential residual ballistic component traveling along with the scattered beam. In the experiments in transmission [Fig. \ref{fig_method_summary}(c), left], the beam exiting the scattering medium is collected at a distance of $\sim$1.5 mm, where a fully developed speckle pattern was observed, with another Olympus PLN20X 0.4 NA objective. The light source (MaiTai HP Ti:Sapphire laser, Spectra-Physics) is set to monochromatic operation mode at a wavelength of 808 nm. The experiments in reflection [Fig. \ref{fig_method_summary}(c), right] reproduce a typical optical coherence tomography (OCT) setup, whereby ultrashort pulses (with a central wavelength of 808 nm and a duration of 100 fs) are sent through the scattering medium and the backscattered, elongated pulses are gated at a time delay set by a delay line along the reference arm.

Dynamics is introduced by transversally translating the scattering medium across to the incident beam, with steps following a two-dimensional random walk with random Gaussian increments, where the standard deviation determines the stability time of the medium.
% Typical values of the average speed range from 20 to 150 nm/s. Such figures should, however, always be compared to the throughput of the whole system, comprising both hardware and software operations.
More details on it will be provided in the next section.

%%%%%%%%%%%%%%%%%%%%%%%%%%%%%%%%%%%%%%%%%%%%%%%%%%%%
%%%%%%%%%%%%%%%%%%  RESULTS  %%%%%%%%%%%%%%%%%%
%%%%%%%%%%%%%%%%%%%%%%%%%%%%%%%%%%%%%%%%%%%%%%%%%%%%
\section{Results}
\label{sec:results}
\begin{figure}[htb]
    \centering\includegraphics[width=\linewidth]{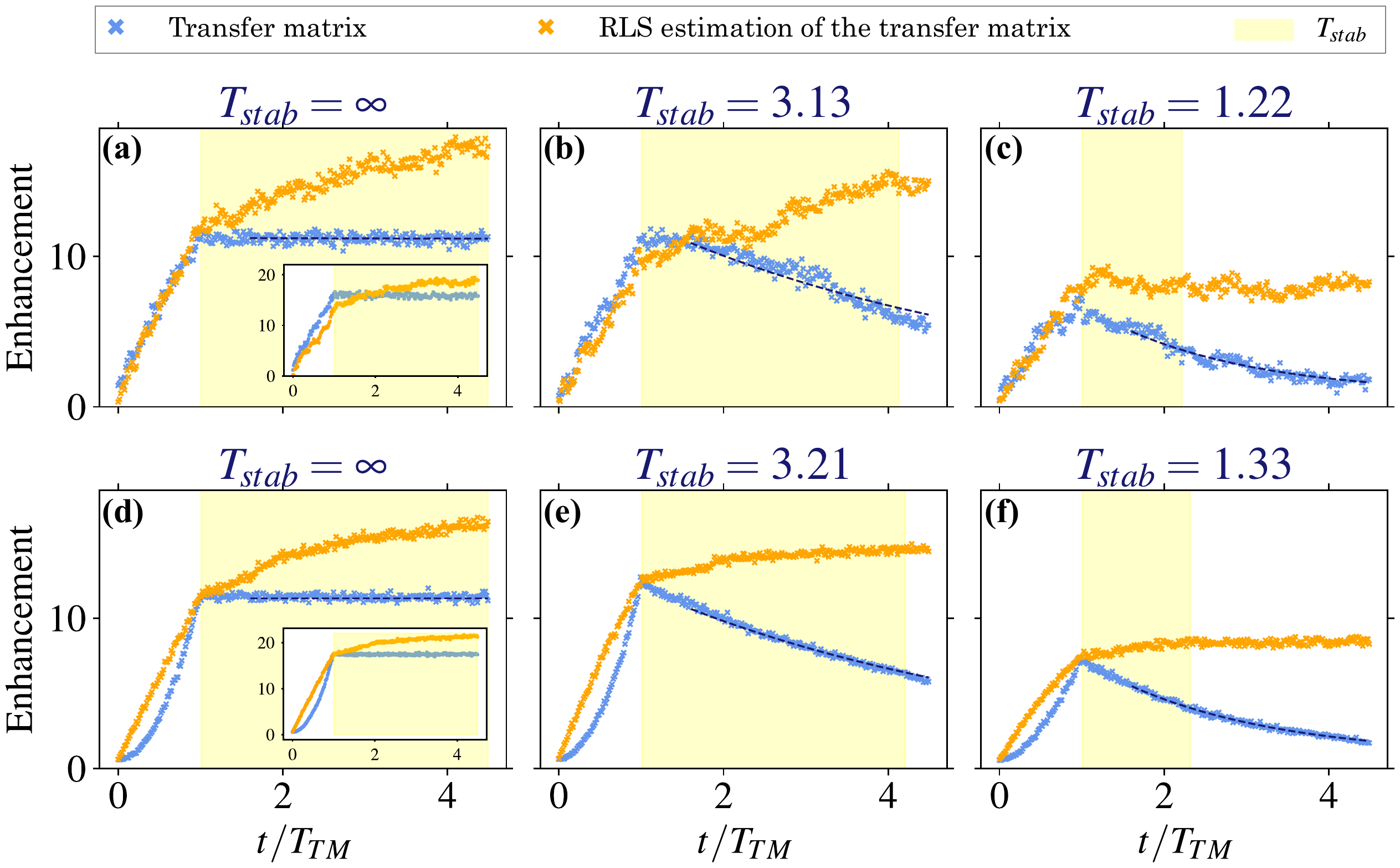}
    \caption{Enhancement of the intensity at one output pixel produced through a scattering medium, as a function time. $T_{TM}$ is the time to optimize over all the $N$ = 64 degrees of freedom once. Blue points: conventional transfer-matrix measurement, lasting $t/T_{TM} = 1$. Its estimate is held constant for later times, allowing the extraction of the stability time $T_{stab}$ of the scattering medium via exponential fitting (dashed dark blue traces). Orange points: RLS estimation of the transfer matrix. (a)-(c): experimental results, averaged over 9 realizations of a focus produced at the center of the camera field of view, upon measuring different regions of the scattering medium. (d)-(f): corresponding simulations at comparable $T_{stab}$. The insets in panels (a), (d) show the same results obtained after doubling the SNR.}
    \label{fig_transmission_with_sims}
\end{figure}

\noindent
Figure \ref{fig_transmission_with_sims} summarizes the performance of the RLS algorithm for the online estimation of the transmission matrix. The beam incident onto the SLM is modulated according to the Hadamard patterns with $N$ = 64 pixels. Every time an input $\boldsymbol{a}_t$ is sent through the scattering medium and the corresponding output field $\boldsymbol{y}_t$ is measured, the inverse QRD-RLS update routine of Algorithm \ref{algo:RLS} is executed, yielding an estimate $\boldsymbol{\hat{X}}_t$ of the transfer matrix. Note, that this procedure can be continuously repeated---after sending the $N$-th input, the first Hadamard vector or any other known input pattern can be sent. As long as the scattering medium is static, probing it with the same input multiple times corresponds to oversampling the unknown $N \times M$ coefficients of its transfer matrix, thereby improving their estimation. It is indeed known that the covariance of the estimated transfer matrix is inversely proportional to $\boldsymbol{C}_t^{-1}$, thus decreasing as $t^{-1}$ \cite{haykin2014adaptive}. Since the true value $\boldsymbol{X}_t$ is unknown, the quality of our reconstruction is evaluated via the intensity of a focus produced behind the scattering medium. We report the intensity enhancement, relative to the average intensity of a non-optimized speckle pattern \cite{vellekoop2015feedback}. The learning curve for a static scattering medium, obtained from the RLS algorithm, is shown as an orange trace in Fig. \ref{fig_transmission_with_sims}(a). The temporal axis is expressed in units of $T_{TM}$, which is defined as the time needed to update the estimation of transfer matrix $N$ times. In other words, a conventional transfer matrix experiments lasts $T_{TM}$. Equivalently, a normalized time of 2 means the oversampling ratio is 2. To showcase the beneficial effect of oversampling, the blue trace shows the performance of a conventional transfer-matrix measurement, using $N$ measurements. At times $t/T_{TM} \leq 1$, the two approaches are equivalent---data are not oversampled. At later times, however, one can take advantage of the whole history of past data to build an estimate more resilient against noise. Our values of the enhancement, when compared to the number of input degrees of freedom $N$, are on a par with previously reported measurements with no oversampling \cite{popoff2010measuring, blochet2017focusing, blochet2019enhanced}.

The same procedure is repeated with dynamic scattering media. By duly tuning the average speed of their movements, we achieve different stability times $T_{stab}$ (also expressed in units of $T_{TM}$). These are estimated as the time constant of an exponential function fitting the tails of the blue traces. At oversampling ratios in the range 3-4, we increase the focus intensity by a factor between 1.5 and 2, compared to the values after a conventional transfer-matrix approach. Upon decreasing $T_{stab}$, the oversampling ratio decreases too, and the performances of the two approaches gradually match, however the RLS estimation always operates in a memory-efficient manner. With dynamic media, forgetting factors $\lambda < 1$ should be used. In our experiments featuring $T_{stab}$ in the range 1-4, we have chosen $1 - \lambda \approx 10^{-5}$, achieving a good compromise between tracking capability and numerical stability. Interestingly, it has been shown that the optimal forgetting factor heavily depends on the number of unknown parameters $N$ which, fortunately, is under user control \cite{ciochina2009influence}. Furthermore, the structure of Algorithm \ref{algo:RLS} suggests that each inverse QRD-RLS iteration may be run at a different value of $\lambda$, allowing the user to pick the one yielding the best performance in an online manner, \textit{i.e.}, with no need to restart the optimization anew and using the current enhancement as a feedback to tune the next value of $\lambda$. Trivially, the optimal value for static media is instead $\lambda = 1$. The best regularization constant $\delta$ depends on the signal-to-noise ratio (SNR) of the measurements. In our experiments, the fact that the RLS algorithm is on a par with the conventional transfer-matrix approach at $t/T_{TM} \leq 1$ and $T_{TM} < T_{stab}$ suggests that the selected regularization constant (here $\delta$ = 1) is optimized for the best performance.

In order to test the validity of \eqref{eq:loss} and further justify our assumptions, the experiments at the top row of Fig. \ref{fig_transmission_with_sims} are reproduced with numerical simulations, following the AR(1) model of \eqref{eq:ar1}  [Figs. \ref{fig_transmission_with_sims}(d)-(f)]. We simulated a finite SNR by corrupting the outputs $\boldsymbol{y}_t$ with additive white complex Gaussian noise with variance $\sigma_{\text{noise}}^2$ and setting SNR $\equiv \sigma_{\boldsymbol{X}}^2 / \sigma_{\text{noise}}^2$. Overall a good quantitative agreement is obtained. For example, if the SNR is increased by a factor of 2 doubling the number of phase-stepped images for field reconstruction, the experimental performance is the one plotted as an inset in Fig. \ref{fig_transmission_with_sims}(a). The same trend is retrieved by simulating measurements with halved $\sigma_{\text{noise}}^2$ [inset in Fig. \ref{fig_transmission_with_sims}(d)].

\begin{figure}[htb]
    \centering\includegraphics[width=\linewidth]{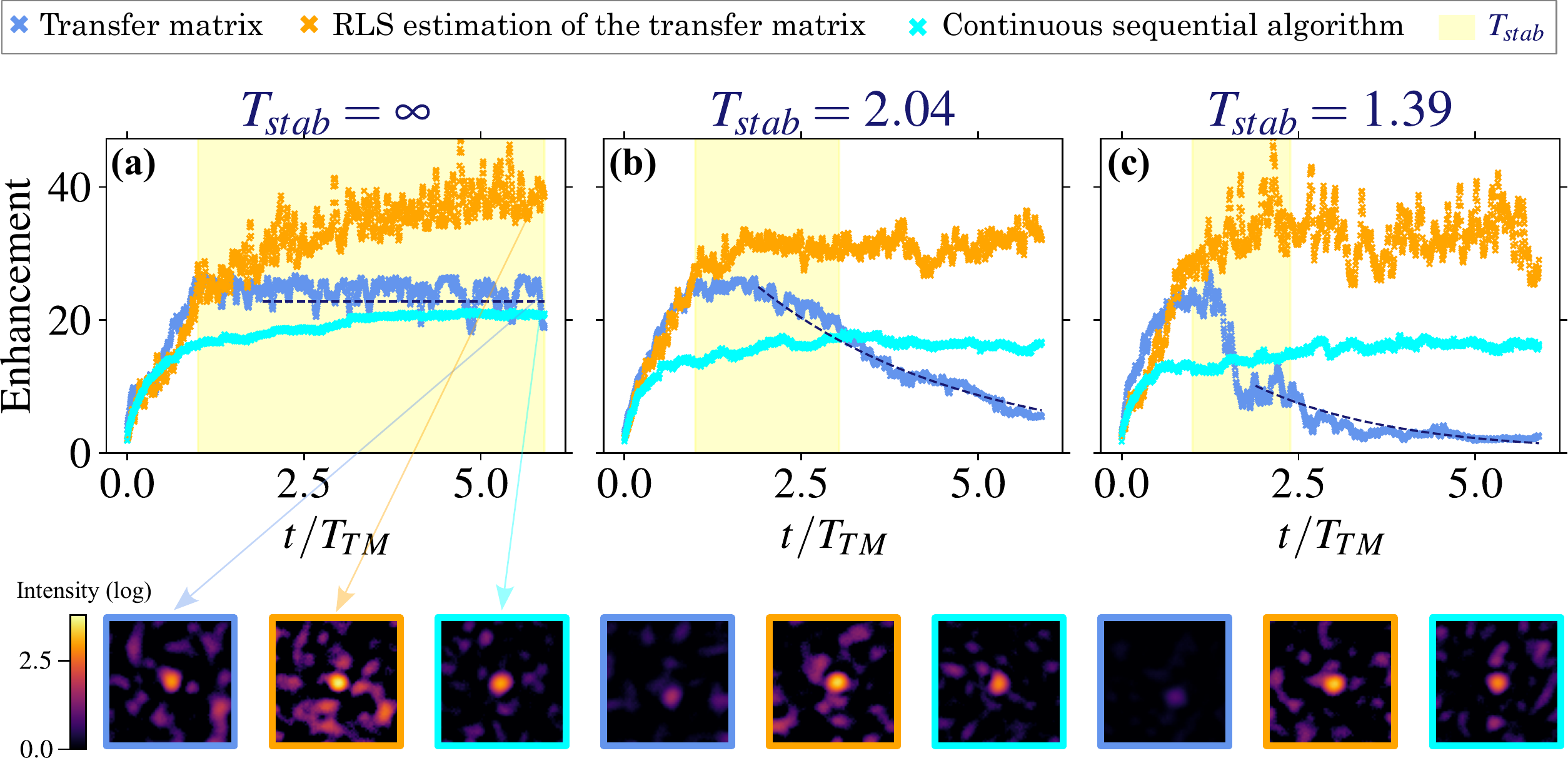}
    \caption{Enhancement of the intensity at one output pixel produced in a scattering medium, as a function time. $T_{TM}$ is the time to optimize over all the $N$ = 256 degrees of freedom once. Blue points: conventional transfer-matrix measurement, lasting $t/T_{TM} = 1$. Its estimate is held constant for later times, allowing the extraction of the stability time $T_{stab}$ of the scattering medium via exponential fitting (dashed dark blue traces). Orange points: RLS estimation of the transfer matrix. Cyan points: optimization via the continuous sequential algorithm (CSA). (a)-(c): experimental tracking performance, averaged over 8 realizations of foci produced across the full camera field of view, hence the higher variability than in Fig. \ref{fig_transmission_with_sims}. The bottom row displays typical images of the focus at the last time step, for each algorithm and each value of $T_{stab}$ (in a common logarithmic scale to ease visibility). Note that these images are not an average of all 8 realizations, but they show one realization only.}
    \label{fig_reflection}
\end{figure}

Analogous results, plotted in Fig. \ref{fig_reflection}, are obtained with the non-invasive OCT setup on the right-hand side of Fig. \ref{fig_method_summary}(c), setting the time delay yielding the maximum average gated intensity. In this instance, the transfer matrix is the time-gated reflection matrix \cite{badon2016smart}. The two learning curves corresponding to the retrieval of the transfer matrix are compared to a conventional optimization routine, which can recursively track the changes in the scattering medium, namely the CSA (in cyan). After blocking the beam along the reference arm, we implement a version of the CSA modulating half of the SLM pixels (corresponding to the +1 or -1 entries of the $N$ Hadamard patterns) at each iteration, yielding the best interference contrast (thus bearing similarities to the partitioning algorithm too \cite{vellekoop2008phase}). It displays comparable performances to a conventional transfer-matrix measurement with a static medium [although convergence is reached later, owing to its stochastic nature, Fig. \ref{fig_reflection}(a)], and it shows solid tracking capabilities in dynamic environments [Figs. \ref{fig_reflection}(b)-(c)]. Still, the resilience to noise of the inverse QRD-RLS algorithm makes it the preferred choice in this setting too, achieving an intensity twice as high as the one obtained with the CSA. The bottom row of Fig. \ref{fig_reflection} shows the corresponding focal spots produced by each algorithm at the last time step.

\begin{figure}[htb]
    \centering\includegraphics[width=\linewidth]{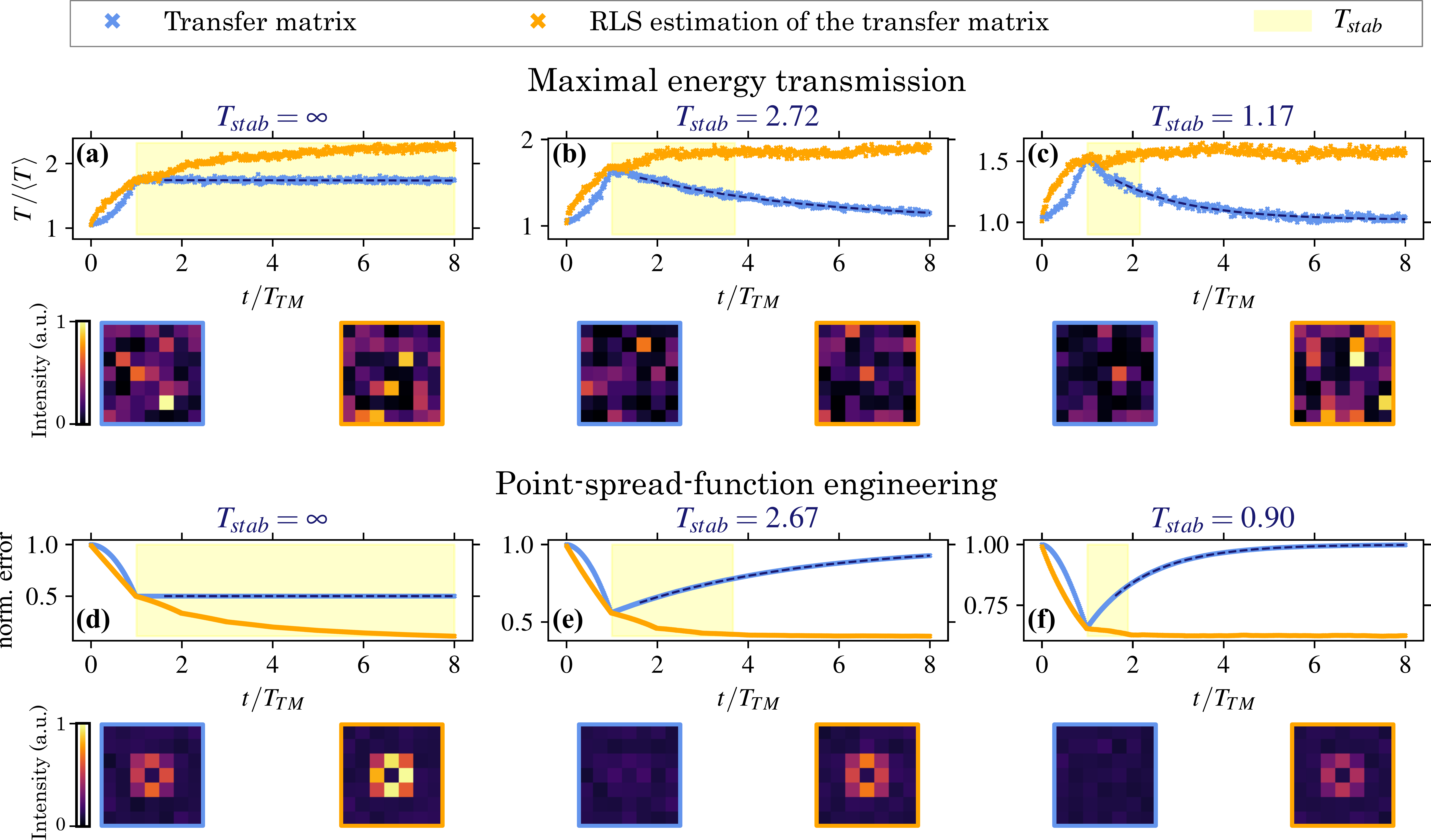}
    \caption{Light control through dynamic scattering media goes beyond focusing. (a)-(c): Enhancement of the total transmittance across the full field of view through a scattering medium, as a function time and for different stability times $T_{stab}$. $T_{TM}$ is the time to optimize over all the $N$ degrees of freedom once. Blue points: conventional transfer-matrix measurement, lasting $t/T_{TM} = 1$. Its estimate is held constant for later times, allowing the extraction of the stability time $T_{stab}$ of the scattering medium via exponential fitting (dashed dark blue traces). Orange points: RLS estimation of the transfer matrix. Here we used $N$ = 64 and $M$ = 49. (d)-(f): normalized error in the estimation of the transfer matrix of the scattering medium, where the same color legend as in (a)-(c) applies. In each set of results, the bottom row displays typical camera images at the last time step. Simulated results.}
    \label{fig_energy_psf}
\end{figure}

Because our proposed routine retrieves the coefficients of the transfer matrix at all output pixels simultaneously, its applications go beyond focusing. Figure \ref{fig_energy_psf} showcases two light-control tasks, through dynamic scattering media, enabled by the recursive and online estimation of the transfer matrix. The first one is maximal energy transmission, upon sending the leading singular vector of the transfer matrix (top row) \cite{kim2012maximal}, and the second one consists in arbitrarily shaping a PSF (bottom row), here in a donut shape \cite{boniface2017transmission}. As expected, these trends replicate the performance on the focusing task of Fig. \ref{fig_transmission_with_sims}.

The most important asset to characterize dynamic scattering media is the wavefront shaping device. Digital micromirror devices (DMD) offer a valuable alternative to liquid-crystal-based arrays and microelectromechanical systems (MEMS) modulators in terms of cost ($\sim$1 kUSD), pixel count ($>$10$^5$) and operating frequencies ($>$10 kHz). Despite their binary amplitude modulation, several strategies have been devised to enable light control through scattering media. Lee holography \cite{lee1974binary, conkey2012high} and superpixel-based related methods \cite{goorden2014superpixel} achieve phase and amplitude control, at the expense of a more involved setup and a relatively low light efficiency, as they rely on an analog spatial filter. Aiming for a simple and non-invasive implementation suitable for real-life applications, Bayesian algorithms have been proposed to solve the phase retrieval problem $\boldsymbol{y}_t = |\boldsymbol{X}_t \boldsymbol{a}_t|^2$ (with $|\cdot|$ denoting the element-wise modulus operation), \textit{i.e.}, recover the transfer matrix from intensity-only measurements and binary amplitude modulation of the inputs, thus transferring the hardware complexity to the software. Examples include the phase retrieval Variational Bayes Expectation-Maximization (prVBEM) \cite{dremeau2015reference, dremeau2015phase} algorithm, the phase retrieval Swept Approximate Message Passing (prSAMP) \cite{rajaei2016intensity} algorithms, and their corresponding compressive version, named phase retrieval Generalized AMP (prGAMP) \cite{metzler2016bm3d}. However, their complexities are of order $\mathcal{O}(t^2)$ per iteration, preventing their application to real-time online learning of long ($t \gg N$) and multidimensional time series.

\begin{figure}[htb]
    \centering\includegraphics[width=\linewidth]{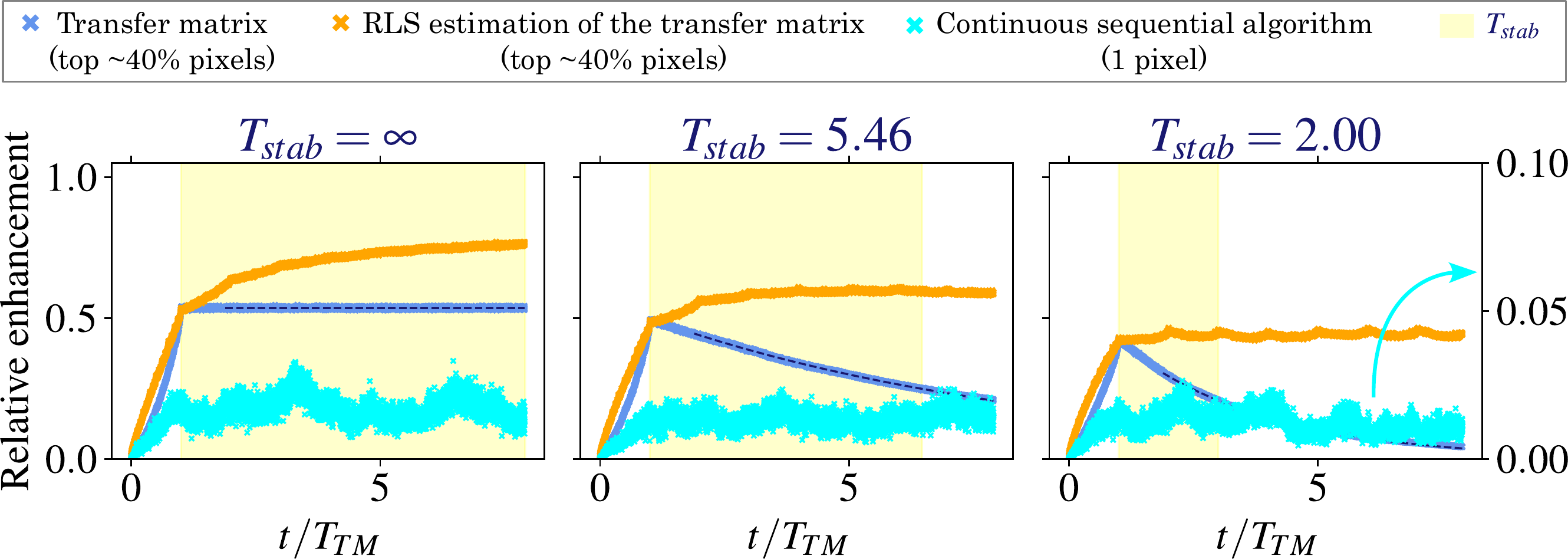}
    \caption{Enhancement of the intensity produced at one output pixel through a scattering medium, as a function time and for different stability times $T_{stab}$. By using the linear approximation of \eqref{eq:binary_mod_linearized} (valid across $\sim$40\% of the output pixels), we achieve wavefront shaping from intensity-only images. $T_{TM}$ is the time to optimize over all the $N$ degrees of freedom once. Blue points: conventional transfer-matrix measurement, lasting $t/T_{TM} = 1$. Its estimate is held constant for later times, allowing the extraction of the stability time $T_{stab}$ of the scattering medium via exponential fitting (dashed dark blue traces). Orange points: RLS estimation of the transfer matrix. Cyan points: continuous sequential algorithm (CSA), plotted on a different scale on the right-hand vertical axis. To reproduce noisy measurements with suboptimal detector performance, all simulated intensities $I$ were corrupted with additive Gaussian noise with a standard deviation of 20$\sqrt{I}$ \cite{yeh2015experimental}. Simulated results.}
    \label{fig_binary_mod}
\end{figure}

In what follows, we show how to implement the RLS estimation technique using non-invasive, intensity-only measurements and binary amplitude modulation of the inputs. When performing wavefront shaping experiments with a DMD, light control is restricted to opening or blocking the modes of the scattering medium, so to achieve the desired output patterns. Hence, the knowledge of the complex-valued transfer matrix is of limited use. We now build on the contribution by Tao and colleagues \cite{tao2015high}. They regard each binary input $\boldsymbol{a}_t \in \{0, 1\}^N$ as the sum of the first Hadamard vector $\boldsymbol{h}_1 = \{1\}^N$, referred to as ``reference'', with any other Hadamard vector $\boldsymbol{h}_t \in \{+1, -1\}^N$, namely $\boldsymbol{a}_t = (\boldsymbol{h}_1 + \boldsymbol{h}_t)/2$. In a similar fashion to inline digital holography, in the output pixels where the reference intensity is larger than the response to an average input, the phase retrieval equation can be linearized. They derive the following linear approximation,
\begin{equation}
    \frac{1}{2} |\boldsymbol{X}_t \boldsymbol{h}_1| \circ \left( |\boldsymbol{X}_t \boldsymbol{a}_t|^2 \oslash |\boldsymbol{X}_t \boldsymbol{h}_1|^2 - \boldsymbol{1} \right) \approx \text{Re} \{ \boldsymbol{X}_t \} \boldsymbol{h}_t \, ,
    \label{eq:binary_mod_linearized}
\end{equation}
where $\circ$ and $\oslash$ denote element-wise vector multiplication and division, respectively, $\boldsymbol{1} \equiv \{1\}^M$ and $\text{Re} \{ \cdot \}$ stands for real part. Note, that the condition for a proper linearization is met, assuming the output pixels are independent, with a probability
\begin{equation}
    \mathcal{P}(I > {\langle I \rangle}) = \int_{{\langle I \rangle}}^{\infty} p(I) dI = e^{-1} \approx 40\% \, ,
\end{equation}
where we have used the probability distribution of the speckle intensity, $p(I) \equiv \exp \left( -I/ \langle I \rangle \right) / {\langle I \rangle}$ \cite{goodman2007speckle}. As all the terms in its left-hand side $\boldsymbol{\tilde{y}}_t$ are known, we can recursively solve \eqref{eq:binary_mod_linearized} for $\text{Re} \{ \boldsymbol{X}_t \}$, minimizing a loss function like the one in \eqref{eq:loss}, and interpreting $\boldsymbol{h}_t$ and $\boldsymbol{\tilde{y}}_t$ as real inputs and outputs, respectively. The real (or, equivalently, imaginary) part of the transfer matrix is all is needed to focus at any output pixel where the linear approximation holds. The corresponding results in Fig. \ref{fig_binary_mod} indeed show the same trend as in Figs. \ref{fig_transmission_with_sims}, \ref{fig_reflection} and \ref{fig_energy_psf}. Here, the enhancement is expressed relative to the maximum enhancement achievable with binary amplitude modulation $\approx 1 + (N/2 - 1)/ \pi$ \cite{akbulut2011focusing}. Feedback-based routines, like the binary version of the CSA \cite{akbulut2011focusing} (plotted in cyan in Fig. \ref{fig_binary_mod}), are highly impacted by experimental noise, as they rely on one single output value. In contrast, exploiting the past data allows us to provide solutions more resilient to noise.

\begin{figure}[htb]
    \centering\includegraphics[width=0.8\linewidth]{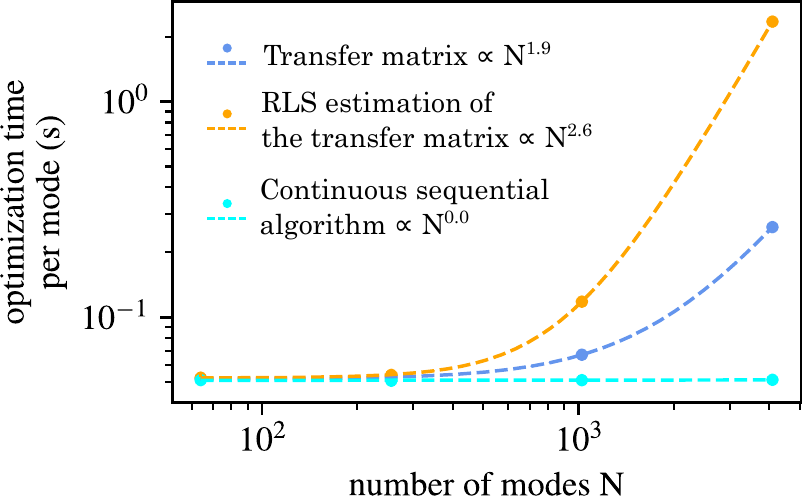}
    \caption{Time required to update the optimal focusing pattern from one new piece of data, estimated in the experimental setup of Fig, \ref{fig_method_summary}(c) right, as a function of the number of input modes. Four phase-stepped intensity images are combined to estimate each output field across a field of view of $M$ = 256 pixels. Blue points and line: conventional transfer-matrix measurement; orange: RLS estimation of the transfer matrix; cyan: continuous sequential algorithm (CSA). Each point is an average of 4096 measurements, such that the standard deviation of the mean is always within the marker size.}
    \label{fig_timing}
\end{figure}

To gain more insight into the performance of our experimental system, its throughput is estimated with the parameters from \cite{popoff2010measuring}, namely $M$ = 256 and 4 phase-shifted intensity images to evaluate each output field. In Fig. \ref{fig_timing} we plot, as a function of the number of input modes $N$, the time to update the optimal focusing pattern from one new piece of data, therefore comprising one (complex) output measurement, the update of the transfer matrix and the computation of the optimal input pattern. For a sufficiently low number of input modes ($N \leq$ 256 in our implementation), the bottleneck is set by the refresh rate of the SLM---we indeed recover a baseline at $\sim$50 ms, which is consistent with the response time $\gtrsim$ 10 ms reported by the manufacturer. With increasing values of $N$, the computation of the optimal pattern takes a non-negligible time at each iteration, hence an onset at $N\sim256$ is observed. In a conventional transfer-matrix measurement (blue line and data points) performed with Hadamard inputs, an additional $\mathcal{O}(N^2)$ is required to bring the optimal focusing pattern from the Hadamard to the canonical basis. The inverse QRD-RLS estimation technique (orange line and data points), based on Algorithm \ref{algo:RLS}, would run with a $\mathcal{O}(N^3)$ complexity, as it involves a QR decomposition \cite{trefethen1997numerical}, but we retrieve a lower power dependence ($\sim$2.6) owing to the low number of data points above the onset. We should, however, recall that Algorithm \ref{algo:RLS} has been implemented to enjoy superior numerical stability. A typical RLS algorithm propagating the inverse covariance matrix instead of its square root would require $N^2$ operations, thus matching the performance of a conventional transfer-matrix measurement. Owing to its updating routine, the iteration time of the CSA (cyan line and data points) is not impacted by the number of input modes, however its performance is limited in dynamic and noisy environments as shown above. As a final remark we stress that online optimization is run on the CPU of an Intel Core i7-6700 processor with 4 cores, a clock speed of 3.4 GHz and 16 GB RAM, thus yielding the onset at $N\sim256$. Therefore our experiments optimize over $N \leq 256$ modes. However, such figures can definitely be increased on a high-performance computing platform.
%

%%%%%%%%%%%%%%%%%%%%%%%%%%%%%%%%%%%%%%%%%%%%%%%%%%%%
%%%%%%%%%%%%%%%%%%  OUTLOOK  %%%%%%%%%%%%%%%%%%
%%%%%%%%%%%%%%%%%%%%%%%%%%%%%%%%%%%%%%%%%%%%%%%%%%%%
\section{Outlook}
\label{sec:outlook}
\noindent
We have presented a recursive and online optimization procedure for the estimation of the transfer matrix of dynamic scattering media, combining the benefits of optimization-based routines and transfer-matrix measurements in wavefront shaping. Experimental and numerical demonstrations have been provided on conventional wavefront shaping setups and for different light-control tasks, noise levels and stability times. Its most intriguing feature is the possibility to optimize multi- and high-dimensional transfer matrices, without the need to store the history of past data in memory. Therefore, we foresee our method to turn out pivotal whenever the scattering behaviour of living biological specimens has to be tracked at various timescales.

\begin{figure}[htb]
    \centering\includegraphics[width=0.9\linewidth]{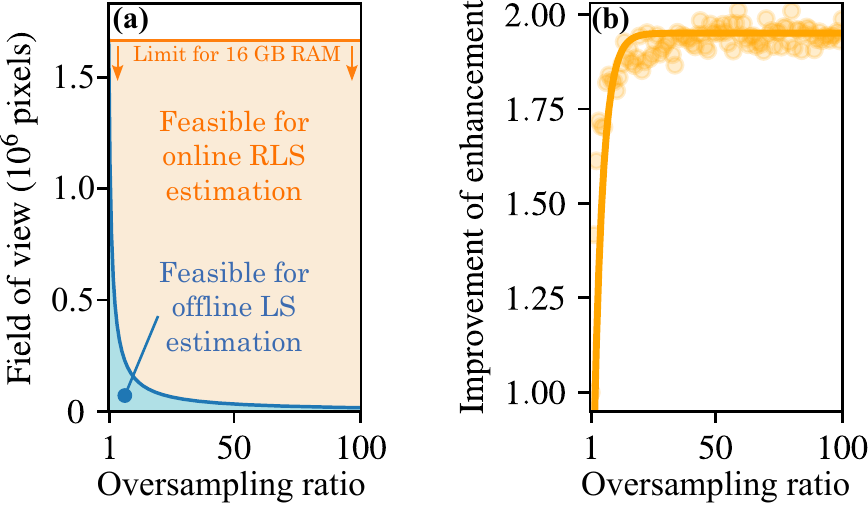}
    \caption{(a) Feasibility regions of the offline least-squares algorithm (blue shaded area) and of the RLS algorithm (orange shaded area), on the plane spanned by the oversampling ratio and the field of view. (b) Evolution of the enhancement as a function of the oversampling ratio (relative to the value at oversampling = 1) at SNR = 1. Points: simulations; line: exponential fit.}
    \label{fig_outlook}
\end{figure}

In our proof-of-principle experiments, all optical modes change with the same rate, therefore they share the same oversampling ratio. However, when imaging large fields of view ($\sim10^4 \si{\micro\meter}^2$) in biological media, timescales differing by factors as large as 100 are accessible. For example, the modes induced by blood flowing decorrelate in less than 10 ms ($>$100 Hz), while breathing modes can last as long as 800 ms (1.25 Hz) in mice \cite{liu2015optical}. As as result of that, the slowest modes enjoy an oversampling ratio close to 100. This means that, compared to an offline least-square estimation of the transfer matrix, a factor of 100 is saved in memory, which can be ultimately used to enlarge the field of view by 2 orders of magnitude. Using the latest MEMS modulators, $N$ = 600 modes can be optimized at a rate of 60 kHz in 10 ms, thus allowing the transfer matrix to be estimated at $M \sim 1.6 \cdot10^6$ output pixels in parallel, assuming 16 GB RAM and double-precision floating-point format (16 B per complex matrix element). This is illustrated in Fig. \ref{fig_outlook}(a), where the feasibility region for offline least-squares is shaded in blue and depends on the oversampling ratio. On the other hand, using the RLS estimation means oversampling does not play a role, so its feasibility region is much larger (orange shaded area). If we also consider that, in ultrafast wavefront shaping systems like \cite{blochet2017focusing}, the SNR approaches 1, at an oversampling ratio of 100 the RLS estimation of the transfer matrix yields an improvement of the focus intensity by a factor of 2, compared to a conventional transfer-matrix measurement with no oversampling [Fig. \ref{fig_outlook}(b)]. Focusing deep inside scattering media, at locations characterized by a specific stability time, may become a reality, thanks to the recent advancements in optimal light control, exploiting the knowledge of the transfer matrix measured at different times \cite{bouchet2021maximum, bouchet2021optimal}.

Besides sharing the same stability times, all the optical modes considered here are also unpredictable, as they feature random and independent increments according to \eqref{eq:ar1}. Should one possess prior knowledge on the medium dynamics (for example, when dealing with the oscillatory movements due to breathing), the RLS estimation may even be employed to predict future scattering behaviours as well as informing the user on the next most informative inputs to optimize information retrieval \cite{rafayelyan2020large}.

We would finally like to remind that the effectiveness of RLS algorithm is enabled by the linear relationship between the input and output patterns. Linearity is guaranteed by light-matter interaction via elastic scattering and thanks to our measurement scheme, allowing quantitative phase estimation of the output fields. When a non-linear transfer function was involved (Fig. \ref{fig_binary_mod}), a linear approximation was made, at the cost of reduced performance. Towards the recursive optimization of non-linear functions, a kernel version of the RLS algorithm has been proposed \cite{engel2004kernel}. It relies on performing linear regressions in a higher ($>N$) dimensional feature space, approximating the non-linear function. Its implementation, although more complicated than its linear counterpart, would be worth investigating, as it would unlock online learning of the transfer matrix of dynamic scattering media for a wide variety of contrast mechanisms, from fluorescence to non-linear coherent scattering.
\section*{Funding}
\noindent
Swiss National Science Foundation (project P400P2\_199329); H2020 European Research Council (SMARTIES-724473); Horizon 2020 Framework Programme (863203).

%%%%%%%%%%%%%%%%%%%%%%%%%%%%%%%%%%%%%%%%%%%%%%%%%%%%
%%%%%%%%%%%%%%%%%%  ACKNOWLEDGMENTS  %%%%%%%%%%%%%%%%%%
%%%%%%%%%%%%%%%%%%%%%%%%%%%%%%%%%%%%%%%%%%%%%%%%%%%%
\section*{Acknowledgments}
\noindent
We wish to thank Fabrice Harms (Imagine Optic) for drawing our attention to RLS algorithms and Bernhard Rauer (LKB), Louisiane Devaud (LKB) and Jonathan Dong (EPFL) for contributing insightful comments.

% \appendix

% \section{Others}

\newpage
%%%%%%%%%%%%%%%%%%%%%%%%%%%%%%%%%%%%%%%%%%%%%%%%%%%%
%%%%%%%%%%%%%%%%%%  BIBLIO  %%%%%%%%%%%%%%%%%%
%%%%%%%%%%%%%%%%%%%%%%%%%%%%%%%%%%%%%%%%%%%%%%%%%%%%
\bibliography{biblio}

%apsrev4-2.bst 2019-01-14 (MD) hand-edited version of apsrev4-1.bst
%Control: key (0)
%Control: author (8) initials jnrlst
%Control: editor formatted (1) identically to author
%Control: production of article title (0) allowed
%Control: page (0) single
%Control: year (1) truncated
%Control: production of eprint (0) enabled
\begin{thebibliography}{44}%
\makeatletter
\providecommand \@ifxundefined [1]{%
 \@ifx{#1\undefined}
}%
\providecommand \@ifnum [1]{%
 \ifnum #1\expandafter \@firstoftwo
 \else \expandafter \@secondoftwo
 \fi
}%
\providecommand \@ifx [1]{%
 \ifx #1\expandafter \@firstoftwo
 \else \expandafter \@secondoftwo
 \fi
}%
\providecommand \natexlab [1]{#1}%
\providecommand \enquote  [1]{``#1''}%
\providecommand \bibnamefont  [1]{#1}%
\providecommand \bibfnamefont [1]{#1}%
\providecommand \citenamefont [1]{#1}%
\providecommand \href@noop [0]{\@secondoftwo}%
\providecommand \href [0]{\begingroup \@sanitize@url \@href}%
\providecommand \@href[1]{\@@startlink{#1}\@@href}%
\providecommand \@@href[1]{\endgroup#1\@@endlink}%
\providecommand \@sanitize@url [0]{\catcode `\\12\catcode `\$12\catcode
  `\&12\catcode `\#12\catcode `\^12\catcode `\_12\catcode `\%12\relax}%
\providecommand \@@startlink[1]{}%
\providecommand \@@endlink[0]{}%
\providecommand \url  [0]{\begingroup\@sanitize@url \@url }%
\providecommand \@url [1]{\endgroup\@href {#1}{\urlprefix }}%
\providecommand \urlprefix  [0]{URL }%
\providecommand \Eprint [0]{\href }%
\providecommand \doibase [0]{https://doi.org/}%
\providecommand \selectlanguage [0]{\@gobble}%
\providecommand \bibinfo  [0]{\@secondoftwo}%
\providecommand \bibfield  [0]{\@secondoftwo}%
\providecommand \translation [1]{[#1]}%
\providecommand \BibitemOpen [0]{}%
\providecommand \bibitemStop [0]{}%
\providecommand \bibitemNoStop [0]{.\EOS\space}%
\providecommand \EOS [0]{\spacefactor3000\relax}%
\providecommand \BibitemShut  [1]{\csname bibitem#1\endcsname}%
\let\auto@bib@innerbib\@empty
%</preamble>
\bibitem [{\citenamefont {Mertz}(2019)}]{mertz2019introduction}%
  \BibitemOpen
  \bibfield  {author} {\bibinfo {author} {\bibfnamefont {J.}~\bibnamefont
  {Mertz}},\ }\href@noop {} {\emph {\bibinfo {title} {Introduction to optical
  microscopy}}}\ (\bibinfo  {publisher} {Cambridge University Press},\ \bibinfo
  {year} {2019})\BibitemShut {NoStop}%
\bibitem [{\citenamefont {Ellis-Davies}(2007)}]{ellis2007caged}%
  \BibitemOpen
  \bibfield  {author} {\bibinfo {author} {\bibfnamefont {G.~C.}\ \bibnamefont
  {Ellis-Davies}},\ }\bibfield  {title} {\bibinfo {title} {Caged compounds:
  photorelease technology for control of cellular chemistry and physiology},\
  }\href {https://www.nature.com/articles/nmeth1072} {\bibfield  {journal}
  {\bibinfo  {journal} {Nat. Methods}\ }\textbf {\bibinfo {volume} {4}},\
  \bibinfo {pages} {619} (\bibinfo {year} {2007})}\BibitemShut {NoStop}%
\bibitem [{\citenamefont {Lichtman}\ and\ \citenamefont
  {Conchello}(2005)}]{lichtman2005fluorescence}%
  \BibitemOpen
  \bibfield  {author} {\bibinfo {author} {\bibfnamefont {J.~W.}\ \bibnamefont
  {Lichtman}}\ and\ \bibinfo {author} {\bibfnamefont {J.-A.}\ \bibnamefont
  {Conchello}},\ }\bibfield  {title} {\bibinfo {title} {Fluorescence
  microscopy},\ }\href {https://www.nature.com/articles/nmeth817} {\bibfield
  {journal} {\bibinfo  {journal} {Nat. Methods}\ }\textbf {\bibinfo {volume}
  {2}},\ \bibinfo {pages} {910} (\bibinfo {year} {2005})}\BibitemShut {NoStop}%
\bibitem [{\citenamefont {Rotter}\ and\ \citenamefont
  {Gigan}(2017)}]{rotter2017light}%
  \BibitemOpen
  \bibfield  {author} {\bibinfo {author} {\bibfnamefont {S.}~\bibnamefont
  {Rotter}}\ and\ \bibinfo {author} {\bibfnamefont {S.}~\bibnamefont {Gigan}},\
  }\bibfield  {title} {\bibinfo {title} {Light fields in complex media:
  Mesoscopic scattering meets wave control},\ }\href
  {https://journals.aps.org/rmp/abstract/10.1103/RevModPhys.89.015005}
  {\bibfield  {journal} {\bibinfo  {journal} {Rev. Mod. Phys.}\ }\textbf
  {\bibinfo {volume} {89}},\ \bibinfo {pages} {015005} (\bibinfo {year}
  {2017})}\BibitemShut {NoStop}%
\bibitem [{\citenamefont {Liu}\ \emph {et~al.}(2015)\citenamefont {Liu},
  \citenamefont {Lai}, \citenamefont {Ma}, \citenamefont {Xu}, \citenamefont
  {Grabar},\ and\ \citenamefont {Wang}}]{liu2015optical}%
  \BibitemOpen
  \bibfield  {author} {\bibinfo {author} {\bibfnamefont {Y.}~\bibnamefont
  {Liu}}, \bibinfo {author} {\bibfnamefont {P.}~\bibnamefont {Lai}}, \bibinfo
  {author} {\bibfnamefont {C.}~\bibnamefont {Ma}}, \bibinfo {author}
  {\bibfnamefont {X.}~\bibnamefont {Xu}}, \bibinfo {author} {\bibfnamefont
  {A.~A.}\ \bibnamefont {Grabar}},\ and\ \bibinfo {author} {\bibfnamefont
  {L.~V.}\ \bibnamefont {Wang}},\ }\bibfield  {title} {\bibinfo {title}
  {Optical focusing deep inside dynamic scattering media with near-infrared
  time-reversed ultrasonically encoded ({TRUE}) light},\ }\href
  {https://www.nature.com/articles/ncomms6904} {\bibfield  {journal} {\bibinfo
  {journal} {Nat. Commun.}\ }\textbf {\bibinfo {volume} {6}},\ \bibinfo {pages}
  {1} (\bibinfo {year} {2015})}\BibitemShut {NoStop}%
\bibitem [{\citenamefont {Assmus}\ \emph {et~al.}(2006)\citenamefont {Assmus},
  \citenamefont {Herwig}, \citenamefont {Cho},\ and\ \citenamefont
  {Wolkenhauer}}]{assmus2006dynamics}%
  \BibitemOpen
  \bibfield  {author} {\bibinfo {author} {\bibfnamefont {H.~E.}\ \bibnamefont
  {Assmus}}, \bibinfo {author} {\bibfnamefont {R.}~\bibnamefont {Herwig}},
  \bibinfo {author} {\bibfnamefont {K.-H.}\ \bibnamefont {Cho}},\ and\ \bibinfo
  {author} {\bibfnamefont {O.}~\bibnamefont {Wolkenhauer}},\ }\bibfield
  {title} {\bibinfo {title} {Dynamics of biological systems: role of systems
  biology in medical research},\ }\href
  {https://www.tandfonline.com/doi/abs/10.1586/14737159.6.6.891} {\bibfield
  {journal} {\bibinfo  {journal} {Expert Rev. Mol. Diagn.}\ }\textbf {\bibinfo
  {volume} {6}},\ \bibinfo {pages} {891} (\bibinfo {year} {2006})}\BibitemShut
  {NoStop}%
\bibitem [{\citenamefont {Engl}\ \emph {et~al.}(2009)\citenamefont {Engl},
  \citenamefont {Flamm}, \citenamefont {K{\"u}gler}, \citenamefont {Lu},
  \citenamefont {M{\"u}ller},\ and\ \citenamefont
  {Schuster}}]{engl2009inverse}%
  \BibitemOpen
  \bibfield  {author} {\bibinfo {author} {\bibfnamefont {H.~W.}\ \bibnamefont
  {Engl}}, \bibinfo {author} {\bibfnamefont {C.}~\bibnamefont {Flamm}},
  \bibinfo {author} {\bibfnamefont {P.}~\bibnamefont {K{\"u}gler}}, \bibinfo
  {author} {\bibfnamefont {J.}~\bibnamefont {Lu}}, \bibinfo {author}
  {\bibfnamefont {S.}~\bibnamefont {M{\"u}ller}},\ and\ \bibinfo {author}
  {\bibfnamefont {P.}~\bibnamefont {Schuster}},\ }\bibfield  {title} {\bibinfo
  {title} {Inverse problems in systems biology},\ }\href
  {https://iopscience.iop.org/article/10.1088/0266-5611/25/12/123014/meta}
  {\bibfield  {journal} {\bibinfo  {journal} {Inverse Probl.}\ }\textbf
  {\bibinfo {volume} {25}},\ \bibinfo {pages} {123014} (\bibinfo {year}
  {2009})}\BibitemShut {NoStop}%
\bibitem [{\citenamefont {Vellekoop}(2015)}]{vellekoop2015feedback}%
  \BibitemOpen
  \bibfield  {author} {\bibinfo {author} {\bibfnamefont {I.~M.}\ \bibnamefont
  {Vellekoop}},\ }\bibfield  {title} {\bibinfo {title} {Feedback-based
  wavefront shaping},\ }\href
  {https://opg.optica.org/oe/fulltext.cfm?uri=oe-23-9-12189&id=315860}
  {\bibfield  {journal} {\bibinfo  {journal} {Opt. Express}\ }\textbf {\bibinfo
  {volume} {23}},\ \bibinfo {pages} {12189} (\bibinfo {year}
  {2015})}\BibitemShut {NoStop}%
\bibitem [{\citenamefont {Devaud}\ \emph {et~al.}(2021)\citenamefont {Devaud},
  \citenamefont {Rauer}, \citenamefont {Melchard}, \citenamefont
  {K{\"u}hmayer}, \citenamefont {Rotter},\ and\ \citenamefont
  {Gigan}}]{devaud2021speckle}%
  \BibitemOpen
  \bibfield  {author} {\bibinfo {author} {\bibfnamefont {L.}~\bibnamefont
  {Devaud}}, \bibinfo {author} {\bibfnamefont {B.}~\bibnamefont {Rauer}},
  \bibinfo {author} {\bibfnamefont {J.}~\bibnamefont {Melchard}}, \bibinfo
  {author} {\bibfnamefont {M.}~\bibnamefont {K{\"u}hmayer}}, \bibinfo {author}
  {\bibfnamefont {S.}~\bibnamefont {Rotter}},\ and\ \bibinfo {author}
  {\bibfnamefont {S.}~\bibnamefont {Gigan}},\ }\bibfield  {title} {\bibinfo
  {title} {Speckle engineering through singular value decomposition of the
  transmission matrix},\ }\href
  {https://journals.aps.org/prl/abstract/10.1103/PhysRevLett.127.093903}
  {\bibfield  {journal} {\bibinfo  {journal} {Phys. Rev. Lett.}\ }\textbf
  {\bibinfo {volume} {127}},\ \bibinfo {pages} {093903} (\bibinfo {year}
  {2021})}\BibitemShut {NoStop}%
\bibitem [{\citenamefont {Boniface}\ \emph {et~al.}(2017)\citenamefont
  {Boniface}, \citenamefont {Mounaix}, \citenamefont {Blochet}, \citenamefont
  {Piestun},\ and\ \citenamefont {Gigan}}]{boniface2017transmission}%
  \BibitemOpen
  \bibfield  {author} {\bibinfo {author} {\bibfnamefont {A.}~\bibnamefont
  {Boniface}}, \bibinfo {author} {\bibfnamefont {M.}~\bibnamefont {Mounaix}},
  \bibinfo {author} {\bibfnamefont {B.}~\bibnamefont {Blochet}}, \bibinfo
  {author} {\bibfnamefont {R.}~\bibnamefont {Piestun}},\ and\ \bibinfo {author}
  {\bibfnamefont {S.}~\bibnamefont {Gigan}},\ }\bibfield  {title} {\bibinfo
  {title} {Transmission-matrix-based point-spread-function engineering through
  a complex medium},\ }\href
  {https://opg.optica.org/optica/fulltext.cfm?uri=optica-4-1-54&id=356978}
  {\bibfield  {journal} {\bibinfo  {journal} {Optica}\ }\textbf {\bibinfo
  {volume} {4}},\ \bibinfo {pages} {54} (\bibinfo {year} {2017})}\BibitemShut
  {NoStop}%
\bibitem [{\citenamefont {Popoff}\ \emph
  {et~al.}(2010{\natexlab{a}})\citenamefont {Popoff}, \citenamefont {Lerosey},
  \citenamefont {Fink}, \citenamefont {Boccara},\ and\ \citenamefont
  {Gigan}}]{popoff2010image}%
  \BibitemOpen
  \bibfield  {author} {\bibinfo {author} {\bibfnamefont {S.}~\bibnamefont
  {Popoff}}, \bibinfo {author} {\bibfnamefont {G.}~\bibnamefont {Lerosey}},
  \bibinfo {author} {\bibfnamefont {M.}~\bibnamefont {Fink}}, \bibinfo {author}
  {\bibfnamefont {A.~C.}\ \bibnamefont {Boccara}},\ and\ \bibinfo {author}
  {\bibfnamefont {S.}~\bibnamefont {Gigan}},\ }\bibfield  {title} {\bibinfo
  {title} {Image transmission through an opaque material},\ }\href
  {https://www.nature.com/articles/ncomms1078} {\bibfield  {journal} {\bibinfo
  {journal} {Nat. Commun.}\ }\textbf {\bibinfo {volume} {1}},\ \bibinfo {pages}
  {81} (\bibinfo {year} {2010}{\natexlab{a}})}\BibitemShut {NoStop}%
\bibitem [{\citenamefont {Kim}\ \emph {et~al.}(2012)\citenamefont {Kim},
  \citenamefont {Choi}, \citenamefont {Yoon}, \citenamefont {Choi},
  \citenamefont {Kim}, \citenamefont {Park},\ and\ \citenamefont
  {Choi}}]{kim2012maximal}%
  \BibitemOpen
  \bibfield  {author} {\bibinfo {author} {\bibfnamefont {M.}~\bibnamefont
  {Kim}}, \bibinfo {author} {\bibfnamefont {Y.}~\bibnamefont {Choi}}, \bibinfo
  {author} {\bibfnamefont {C.}~\bibnamefont {Yoon}}, \bibinfo {author}
  {\bibfnamefont {W.}~\bibnamefont {Choi}}, \bibinfo {author} {\bibfnamefont
  {J.}~\bibnamefont {Kim}}, \bibinfo {author} {\bibfnamefont {Q.-H.}\
  \bibnamefont {Park}},\ and\ \bibinfo {author} {\bibfnamefont
  {W.}~\bibnamefont {Choi}},\ }\bibfield  {title} {\bibinfo {title} {Maximal
  energy transport through disordered media with the implementation of
  transmission eigenchannels},\ }\href
  {https://www.nature.com/articles/nphoton.2012.159} {\bibfield  {journal}
  {\bibinfo  {journal} {Nat. Photonics}\ }\textbf {\bibinfo {volume} {6}},\
  \bibinfo {pages} {581} (\bibinfo {year} {2012})}\BibitemShut {NoStop}%
\bibitem [{\citenamefont {Popoff}\ \emph
  {et~al.}(2010{\natexlab{b}})\citenamefont {Popoff}, \citenamefont {Lerosey},
  \citenamefont {Carminati}, \citenamefont {Fink}, \citenamefont {Boccara},\
  and\ \citenamefont {Gigan}}]{popoff2010measuring}%
  \BibitemOpen
  \bibfield  {author} {\bibinfo {author} {\bibfnamefont {S.}~\bibnamefont
  {Popoff}}, \bibinfo {author} {\bibfnamefont {G.}~\bibnamefont {Lerosey}},
  \bibinfo {author} {\bibfnamefont {R.}~\bibnamefont {Carminati}}, \bibinfo
  {author} {\bibfnamefont {M.}~\bibnamefont {Fink}}, \bibinfo {author}
  {\bibfnamefont {A.}~\bibnamefont {Boccara}},\ and\ \bibinfo {author}
  {\bibfnamefont {S.}~\bibnamefont {Gigan}},\ }\bibfield  {title} {\bibinfo
  {title} {Measuring the transmission matrix in optics: an approach to the
  study and control of light propagation in disordered media},\ }\href
  {https://journals.aps.org/prl/abstract/10.1103/PhysRevLett.104.100601}
  {\bibfield  {journal} {\bibinfo  {journal} {Phys. Rev. Lett.}\ }\textbf
  {\bibinfo {volume} {104}},\ \bibinfo {pages} {100601} (\bibinfo {year}
  {2010}{\natexlab{b}})}\BibitemShut {NoStop}%
\bibitem [{\citenamefont {Matth{\`e}s}\ \emph {et~al.}(2021)\citenamefont
  {Matth{\`e}s}, \citenamefont {Bromberg}, \citenamefont {de~Rosny},\ and\
  \citenamefont {Popoff}}]{matthes2021learning}%
  \BibitemOpen
  \bibfield  {author} {\bibinfo {author} {\bibfnamefont {M.~W.}\ \bibnamefont
  {Matth{\`e}s}}, \bibinfo {author} {\bibfnamefont {Y.}~\bibnamefont
  {Bromberg}}, \bibinfo {author} {\bibfnamefont {J.}~\bibnamefont {de~Rosny}},\
  and\ \bibinfo {author} {\bibfnamefont {S.~M.}\ \bibnamefont {Popoff}},\
  }\bibfield  {title} {\bibinfo {title} {Learning and avoiding disorder in
  multimode fibers},\ }\href
  {https://journals.aps.org/prx/abstract/10.1103/PhysRevX.11.021060} {\bibfield
   {journal} {\bibinfo  {journal} {Phys. Rev. X}\ }\textbf {\bibinfo {volume}
  {11}},\ \bibinfo {pages} {021060} (\bibinfo {year} {2021})}\BibitemShut
  {NoStop}%
\bibitem [{\citenamefont {Conkey}\ \emph
  {et~al.}(2012{\natexlab{a}})\citenamefont {Conkey}, \citenamefont {Brown},
  \citenamefont {Caravaca-Aguirre},\ and\ \citenamefont
  {Piestun}}]{conkey2012genetic}%
  \BibitemOpen
  \bibfield  {author} {\bibinfo {author} {\bibfnamefont {D.~B.}\ \bibnamefont
  {Conkey}}, \bibinfo {author} {\bibfnamefont {A.~N.}\ \bibnamefont {Brown}},
  \bibinfo {author} {\bibfnamefont {A.~M.}\ \bibnamefont {Caravaca-Aguirre}},\
  and\ \bibinfo {author} {\bibfnamefont {R.}~\bibnamefont {Piestun}},\
  }\bibfield  {title} {\bibinfo {title} {Genetic algorithm optimization for
  focusing through turbid media in noisy environments},\ }\href
  {https://opg.optica.org/oe/fulltext.cfm?uri=oe-20-5-4840&id=227654}
  {\bibfield  {journal} {\bibinfo  {journal} {Opt. Express}\ }\textbf {\bibinfo
  {volume} {20}},\ \bibinfo {pages} {4840} (\bibinfo {year}
  {2012}{\natexlab{a}})}\BibitemShut {NoStop}%
\bibitem [{\citenamefont {Cheng}\ \emph {et~al.}(2022)\citenamefont {Cheng},
  \citenamefont {Zhong}, \citenamefont {Woo}, \citenamefont {Zhao},
  \citenamefont {Hui},\ and\ \citenamefont {Lai}}]{cheng2022long}%
  \BibitemOpen
  \bibfield  {author} {\bibinfo {author} {\bibfnamefont {S.}~\bibnamefont
  {Cheng}}, \bibinfo {author} {\bibfnamefont {T.}~\bibnamefont {Zhong}},
  \bibinfo {author} {\bibfnamefont {C.~M.}\ \bibnamefont {Woo}}, \bibinfo
  {author} {\bibfnamefont {Q.}~\bibnamefont {Zhao}}, \bibinfo {author}
  {\bibfnamefont {H.}~\bibnamefont {Hui}},\ and\ \bibinfo {author}
  {\bibfnamefont {P.}~\bibnamefont {Lai}},\ }\bibfield  {title} {\bibinfo
  {title} {Long-distance pattern projection through an unfixed multimode fiber
  with natural evolution strategy-based wavefront shaping},\ }\href
  {https://doi.org/10.1364/OE.462275} {\bibfield  {journal} {\bibinfo
  {journal} {Opt. Express}\ }\textbf {\bibinfo {volume} {30}},\ \bibinfo
  {pages} {32565} (\bibinfo {year} {2022})}\BibitemShut {NoStop}%
\bibitem [{\citenamefont {Haykin}(2014)}]{haykin2014adaptive}%
  \BibitemOpen
  \bibfield  {author} {\bibinfo {author} {\bibfnamefont {S.~S.}\ \bibnamefont
  {Haykin}},\ }\href@noop {} {\emph {\bibinfo {title} {Adaptive filter
  theory}}}\ (\bibinfo  {publisher} {Pearson Education},\ \bibinfo {year}
  {2014})\BibitemShut {NoStop}%
\bibitem [{\citenamefont {Haykin}\ \emph {et~al.}(1997)\citenamefont {Haykin},
  \citenamefont {Sayed}, \citenamefont {Zeidler}, \citenamefont {Yee},\ and\
  \citenamefont {Wei}}]{haykin1997adaptive}%
  \BibitemOpen
  \bibfield  {author} {\bibinfo {author} {\bibfnamefont {S.}~\bibnamefont
  {Haykin}}, \bibinfo {author} {\bibfnamefont {A.~H.}\ \bibnamefont {Sayed}},
  \bibinfo {author} {\bibfnamefont {J.~R.}\ \bibnamefont {Zeidler}}, \bibinfo
  {author} {\bibfnamefont {P.}~\bibnamefont {Yee}},\ and\ \bibinfo {author}
  {\bibfnamefont {P.~C.}\ \bibnamefont {Wei}},\ }\bibfield  {title} {\bibinfo
  {title} {Adaptive tracking of linear time-variant systems by extended {RLS}
  algorithms},\ }\href {https://ieeexplore.ieee.org/abstract/document/575687}
  {\bibfield  {journal} {\bibinfo  {journal} {IEEE Trans. Signal Process.}\
  }\textbf {\bibinfo {volume} {45}},\ \bibinfo {pages} {1118} (\bibinfo {year}
  {1997})}\BibitemShut {NoStop}%
\bibitem [{\citenamefont {Patra}\ \emph {et~al.}(2017)\citenamefont {Patra},
  \citenamefont {Das}, \citenamefont {Mishra},\ and\ \citenamefont
  {Senapati}}]{patra2017adaptive}%
  \BibitemOpen
  \bibfield  {author} {\bibinfo {author} {\bibfnamefont {A.}~\bibnamefont
  {Patra}}, \bibinfo {author} {\bibfnamefont {S.}~\bibnamefont {Das}}, \bibinfo
  {author} {\bibfnamefont {S.}~\bibnamefont {Mishra}},\ and\ \bibinfo {author}
  {\bibfnamefont {M.~R.}\ \bibnamefont {Senapati}},\ }\bibfield  {title}
  {\bibinfo {title} {An adaptive local linear optimized radial basis functional
  neural network model for financial time series prediction},\ }\href
  {https://link.springer.com/article/10.1007/s00521-015-2039-0} {\bibfield
  {journal} {\bibinfo  {journal} {Neural. Comput. Appl.}\ }\textbf {\bibinfo
  {volume} {28}},\ \bibinfo {pages} {101} (\bibinfo {year} {2017})}\BibitemShut
  {NoStop}%
\bibitem [{\citenamefont {Brake}\ \emph {et~al.}(2016)\citenamefont {Brake},
  \citenamefont {Jang},\ and\ \citenamefont {Yang}}]{brake2016analyzing}%
  \BibitemOpen
  \bibfield  {author} {\bibinfo {author} {\bibfnamefont {J.}~\bibnamefont
  {Brake}}, \bibinfo {author} {\bibfnamefont {M.}~\bibnamefont {Jang}},\ and\
  \bibinfo {author} {\bibfnamefont {C.}~\bibnamefont {Yang}},\ }\bibfield
  {title} {\bibinfo {title} {Analyzing the relationship between decorrelation
  time and tissue thickness in acute rat brain slices using multispeckle
  diffusing wave spectroscopy},\ }\href
  {https://opg.optica.org/viewmedia.cfm?r=1&rwjcode=josaa&uri=josaa-33-2-270&html=true}
  {\bibfield  {journal} {\bibinfo  {journal} {J. Opt. Soc. Am. A}\ }\textbf
  {\bibinfo {volume} {33}},\ \bibinfo {pages} {270} (\bibinfo {year}
  {2016})}\BibitemShut {NoStop}%
\bibitem [{\citenamefont {Vellekoop}\ and\ \citenamefont
  {Mosk}(2008)}]{vellekoop2008phase}%
  \BibitemOpen
  \bibfield  {author} {\bibinfo {author} {\bibfnamefont {I.~M.}\ \bibnamefont
  {Vellekoop}}\ and\ \bibinfo {author} {\bibfnamefont {A.}~\bibnamefont
  {Mosk}},\ }\bibfield  {title} {\bibinfo {title} {Phase control algorithms for
  focusing light through turbid media},\ }\href
  {https://www.sciencedirect.com/science/article/abs/pii/S0030401808001430}
  {\bibfield  {journal} {\bibinfo  {journal} {Opt. Commun.}\ }\textbf {\bibinfo
  {volume} {281}},\ \bibinfo {pages} {3071} (\bibinfo {year}
  {2008})}\BibitemShut {NoStop}%
\bibitem [{\citenamefont {Webster}\ \emph {et~al.}(2004)\citenamefont
  {Webster}, \citenamefont {Gerke}, \citenamefont {Weiner},\ and\ \citenamefont
  {Webb}}]{webster2004spectral}%
  \BibitemOpen
  \bibfield  {author} {\bibinfo {author} {\bibfnamefont {M.}~\bibnamefont
  {Webster}}, \bibinfo {author} {\bibfnamefont {T.}~\bibnamefont {Gerke}},
  \bibinfo {author} {\bibfnamefont {A.}~\bibnamefont {Weiner}},\ and\ \bibinfo
  {author} {\bibfnamefont {K.}~\bibnamefont {Webb}},\ }\bibfield  {title}
  {\bibinfo {title} {Spectral and temporal speckle field measurements of a
  random medium},\ }\href
  {https://opg.optica.org/ol/abstract.cfm?uri=ol-29-13-1491} {\bibfield
  {journal} {\bibinfo  {journal} {Opt. Lett.}\ }\textbf {\bibinfo {volume}
  {29}},\ \bibinfo {pages} {1491} (\bibinfo {year} {2004})}\BibitemShut
  {NoStop}%
\bibitem [{\citenamefont {Alexander}\ and\ \citenamefont
  {Ghimikar}(1993)}]{alexander1993method}%
  \BibitemOpen
  \bibfield  {author} {\bibinfo {author} {\bibfnamefont {S.~T.}\ \bibnamefont
  {Alexander}}\ and\ \bibinfo {author} {\bibfnamefont {A.~L.}\ \bibnamefont
  {Ghimikar}},\ }\bibfield  {title} {\bibinfo {title} {A method for recursive
  least squares filtering based upon an inverse {QR} decomposition},\ }\href
  {https://ieeexplore.ieee.org/document/193124} {\bibfield  {journal} {\bibinfo
   {journal} {IEEE Trans. Signal Process.}\ }\textbf {\bibinfo {volume} {41}},\
  \bibinfo {pages} {20} (\bibinfo {year} {1993})}\BibitemShut {NoStop}%
\bibitem [{\citenamefont {Valzania}(2022)}]{github_code}%
  \BibitemOpen
  \bibfield  {author} {\bibinfo {author} {\bibfnamefont {L.}~\bibnamefont
  {Valzania}},\ }\href@noop {} {\bibinfo {title} {Online learning of the
  transfer matrix of dynamic scattering media}},\ \bibinfo {howpublished}
  {\url{https://github.com/laboGigan/online_learning_TM}} (\bibinfo {year}
  {2022})\BibitemShut {NoStop}%
\bibitem [{\citenamefont {Blochet}\ \emph {et~al.}(2017)\citenamefont
  {Blochet}, \citenamefont {Bourdieu},\ and\ \citenamefont
  {Gigan}}]{blochet2017focusing}%
  \BibitemOpen
  \bibfield  {author} {\bibinfo {author} {\bibfnamefont {B.}~\bibnamefont
  {Blochet}}, \bibinfo {author} {\bibfnamefont {L.}~\bibnamefont {Bourdieu}},\
  and\ \bibinfo {author} {\bibfnamefont {S.}~\bibnamefont {Gigan}},\ }\bibfield
   {title} {\bibinfo {title} {Focusing light through dynamical samples using
  fast continuous wavefront optimization},\ }\href
  {https://opg.optica.org/ol/abstract.cfm?uri=ol-42-23-4994} {\bibfield
  {journal} {\bibinfo  {journal} {Opt. Lett.}\ }\textbf {\bibinfo {volume}
  {42}},\ \bibinfo {pages} {4994} (\bibinfo {year} {2017})}\BibitemShut
  {NoStop}%
\bibitem [{\citenamefont {Blochet}\ \emph {et~al.}(2019)\citenamefont
  {Blochet}, \citenamefont {Joaquina}, \citenamefont {Blum}, \citenamefont
  {Bourdieu},\ and\ \citenamefont {Gigan}}]{blochet2019enhanced}%
  \BibitemOpen
  \bibfield  {author} {\bibinfo {author} {\bibfnamefont {B.}~\bibnamefont
  {Blochet}}, \bibinfo {author} {\bibfnamefont {K.}~\bibnamefont {Joaquina}},
  \bibinfo {author} {\bibfnamefont {L.}~\bibnamefont {Blum}}, \bibinfo {author}
  {\bibfnamefont {L.}~\bibnamefont {Bourdieu}},\ and\ \bibinfo {author}
  {\bibfnamefont {S.}~\bibnamefont {Gigan}},\ }\bibfield  {title} {\bibinfo
  {title} {Enhanced stability of the focus obtained by wavefront optimization
  in dynamical scattering media},\ }\href
  {https://opg.optica.org/optica/fulltext.cfm?uri=optica-6-12-1554&id=424745}
  {\bibfield  {journal} {\bibinfo  {journal} {Optica}\ }\textbf {\bibinfo
  {volume} {6}},\ \bibinfo {pages} {1554} (\bibinfo {year} {2019})}\BibitemShut
  {NoStop}%
\bibitem [{\citenamefont {Ciochina}\ \emph {et~al.}(2009)\citenamefont
  {Ciochina}, \citenamefont {Paleologu}, \citenamefont {Benesty},\ and\
  \citenamefont {Enescu}}]{ciochina2009influence}%
  \BibitemOpen
  \bibfield  {author} {\bibinfo {author} {\bibfnamefont {S.}~\bibnamefont
  {Ciochina}}, \bibinfo {author} {\bibfnamefont {C.}~\bibnamefont {Paleologu}},
  \bibinfo {author} {\bibfnamefont {J.}~\bibnamefont {Benesty}},\ and\ \bibinfo
  {author} {\bibfnamefont {A.~A.}\ \bibnamefont {Enescu}},\ }\bibfield  {title}
  {\bibinfo {title} {On the influence of the forgetting factor of the {RLS}
  adaptive filter in system identification},\ }in\ \href
  {https://ieeexplore.ieee.org/abstract/document/5206117} {\emph {\bibinfo
  {booktitle} {2009 International Symposium on Signals, Circuits and
  Systems}}}\ (\bibinfo {organization} {IEEE},\ \bibinfo {year} {2009})\ pp.\
  \bibinfo {pages} {1--4}\BibitemShut {NoStop}%
\bibitem [{\citenamefont {Badon}\ \emph {et~al.}(2016)\citenamefont {Badon},
  \citenamefont {Li}, \citenamefont {Lerosey}, \citenamefont {Boccara},
  \citenamefont {Fink},\ and\ \citenamefont {Aubry}}]{badon2016smart}%
  \BibitemOpen
  \bibfield  {author} {\bibinfo {author} {\bibfnamefont {A.}~\bibnamefont
  {Badon}}, \bibinfo {author} {\bibfnamefont {D.}~\bibnamefont {Li}}, \bibinfo
  {author} {\bibfnamefont {G.}~\bibnamefont {Lerosey}}, \bibinfo {author}
  {\bibfnamefont {A.~C.}\ \bibnamefont {Boccara}}, \bibinfo {author}
  {\bibfnamefont {M.}~\bibnamefont {Fink}},\ and\ \bibinfo {author}
  {\bibfnamefont {A.}~\bibnamefont {Aubry}},\ }\bibfield  {title} {\bibinfo
  {title} {Smart optical coherence tomography for ultra-deep imaging through
  highly scattering media},\ }\href
  {https://www.science.org/doi/10.1126/sciadv.1600370} {\bibfield  {journal}
  {\bibinfo  {journal} {Sci. Adv.}\ }\textbf {\bibinfo {volume} {2}},\ \bibinfo
  {pages} {e1600370} (\bibinfo {year} {2016})}\BibitemShut {NoStop}%
\bibitem [{\citenamefont {Lee}(1974)}]{lee1974binary}%
  \BibitemOpen
  \bibfield  {author} {\bibinfo {author} {\bibfnamefont {W.-H.}\ \bibnamefont
  {Lee}},\ }\bibfield  {title} {\bibinfo {title} {Binary synthetic holograms},\
  }\href {https://opg.optica.org/ao/abstract.cfm?uri=ao-13-7-1677} {\bibfield
  {journal} {\bibinfo  {journal} {Appl. Opt.}\ }\textbf {\bibinfo {volume}
  {13}},\ \bibinfo {pages} {1677} (\bibinfo {year} {1974})}\BibitemShut
  {NoStop}%
\bibitem [{\citenamefont {Conkey}\ \emph
  {et~al.}(2012{\natexlab{b}})\citenamefont {Conkey}, \citenamefont
  {Caravaca-Aguirre},\ and\ \citenamefont {Piestun}}]{conkey2012high}%
  \BibitemOpen
  \bibfield  {author} {\bibinfo {author} {\bibfnamefont {D.~B.}\ \bibnamefont
  {Conkey}}, \bibinfo {author} {\bibfnamefont {A.~M.}\ \bibnamefont
  {Caravaca-Aguirre}},\ and\ \bibinfo {author} {\bibfnamefont {R.}~\bibnamefont
  {Piestun}},\ }\bibfield  {title} {\bibinfo {title} {High-speed scattering
  medium characterization with application to focusing light through turbid
  media},\ }\href
  {https://opg.optica.org/oe/fulltext.cfm?uri=oe-20-2-1733&id=226428}
  {\bibfield  {journal} {\bibinfo  {journal} {Opt. Express}\ }\textbf {\bibinfo
  {volume} {20}},\ \bibinfo {pages} {1733} (\bibinfo {year}
  {2012}{\natexlab{b}})}\BibitemShut {NoStop}%
\bibitem [{\citenamefont {Goorden}\ \emph {et~al.}(2014)\citenamefont
  {Goorden}, \citenamefont {Bertolotti},\ and\ \citenamefont
  {Mosk}}]{goorden2014superpixel}%
  \BibitemOpen
  \bibfield  {author} {\bibinfo {author} {\bibfnamefont {S.~A.}\ \bibnamefont
  {Goorden}}, \bibinfo {author} {\bibfnamefont {J.}~\bibnamefont
  {Bertolotti}},\ and\ \bibinfo {author} {\bibfnamefont {A.~P.}\ \bibnamefont
  {Mosk}},\ }\bibfield  {title} {\bibinfo {title} {Superpixel-based spatial
  amplitude and phase modulation using a digital micromirror device},\ }\href
  {https://opg.optica.org/oe/fulltext.cfm?uri=oe-22-15-17999&id=296147}
  {\bibfield  {journal} {\bibinfo  {journal} {Opt. Express}\ }\textbf {\bibinfo
  {volume} {22}},\ \bibinfo {pages} {17999} (\bibinfo {year}
  {2014})}\BibitemShut {NoStop}%
\bibitem [{\citenamefont {Dr{\'e}meau}\ \emph {et~al.}(2015)\citenamefont
  {Dr{\'e}meau}, \citenamefont {Liutkus}, \citenamefont {Martina},
  \citenamefont {Katz}, \citenamefont {Sch{\"u}lke}, \citenamefont {Krzakala},
  \citenamefont {Gigan},\ and\ \citenamefont {Daudet}}]{dremeau2015reference}%
  \BibitemOpen
  \bibfield  {author} {\bibinfo {author} {\bibfnamefont {A.}~\bibnamefont
  {Dr{\'e}meau}}, \bibinfo {author} {\bibfnamefont {A.}~\bibnamefont
  {Liutkus}}, \bibinfo {author} {\bibfnamefont {D.}~\bibnamefont {Martina}},
  \bibinfo {author} {\bibfnamefont {O.}~\bibnamefont {Katz}}, \bibinfo {author}
  {\bibfnamefont {C.}~\bibnamefont {Sch{\"u}lke}}, \bibinfo {author}
  {\bibfnamefont {F.}~\bibnamefont {Krzakala}}, \bibinfo {author}
  {\bibfnamefont {S.}~\bibnamefont {Gigan}},\ and\ \bibinfo {author}
  {\bibfnamefont {L.}~\bibnamefont {Daudet}},\ }\bibfield  {title} {\bibinfo
  {title} {Reference-less measurement of the transmission matrix of a highly
  scattering material using a {DMD} and phase retrieval techniques},\ }\href
  {https://opg.optica.org/oe/fulltext.cfm?uri=oe-23-9-11898&id=315789}
  {\bibfield  {journal} {\bibinfo  {journal} {Opt. Express}\ }\textbf {\bibinfo
  {volume} {23}},\ \bibinfo {pages} {11898} (\bibinfo {year}
  {2015})}\BibitemShut {NoStop}%
\bibitem [{\citenamefont {Dr{\'e}meau}\ and\ \citenamefont
  {Krzakala}(2015)}]{dremeau2015phase}%
  \BibitemOpen
  \bibfield  {author} {\bibinfo {author} {\bibfnamefont {A.}~\bibnamefont
  {Dr{\'e}meau}}\ and\ \bibinfo {author} {\bibfnamefont {F.}~\bibnamefont
  {Krzakala}},\ }\bibfield  {title} {\bibinfo {title} {Phase recovery from a
  bayesian point of view: the variational approach},\ }in\ \href
  {https://ieeexplore.ieee.org/abstract/document/7178654} {\emph {\bibinfo
  {booktitle} {2015 IEEE International Conference on Acoustics, Speech and
  Signal Processing (ICASSP)}}}\ (\bibinfo {organization} {IEEE},\ \bibinfo
  {year} {2015})\ pp.\ \bibinfo {pages} {3661--3665}\BibitemShut {NoStop}%
\bibitem [{\citenamefont {Rajaei}\ \emph {et~al.}(2016)\citenamefont {Rajaei},
  \citenamefont {Tramel}, \citenamefont {Gigan}, \citenamefont {Krzakala},\
  and\ \citenamefont {Daudet}}]{rajaei2016intensity}%
  \BibitemOpen
  \bibfield  {author} {\bibinfo {author} {\bibfnamefont {B.}~\bibnamefont
  {Rajaei}}, \bibinfo {author} {\bibfnamefont {E.~W.}\ \bibnamefont {Tramel}},
  \bibinfo {author} {\bibfnamefont {S.}~\bibnamefont {Gigan}}, \bibinfo
  {author} {\bibfnamefont {F.}~\bibnamefont {Krzakala}},\ and\ \bibinfo
  {author} {\bibfnamefont {L.}~\bibnamefont {Daudet}},\ }\bibfield  {title}
  {\bibinfo {title} {Intensity-only optical compressive imaging using a
  multiply scattering material and a double phase retrieval approach},\ }in\
  \href {https://ieeexplore.ieee.org/abstract/document/7472439} {\emph
  {\bibinfo {booktitle} {2016 IEEE International Conference on Acoustics,
  Speech and Signal Processing (ICASSP)}}}\ (\bibinfo {organization} {IEEE},\
  \bibinfo {year} {2016})\ pp.\ \bibinfo {pages} {4054--4058}\BibitemShut
  {NoStop}%
\bibitem [{\citenamefont {Metzler}\ \emph {et~al.}(2016)\citenamefont
  {Metzler}, \citenamefont {Maleki},\ and\ \citenamefont
  {Baraniuk}}]{metzler2016bm3d}%
  \BibitemOpen
  \bibfield  {author} {\bibinfo {author} {\bibfnamefont {C.~A.}\ \bibnamefont
  {Metzler}}, \bibinfo {author} {\bibfnamefont {A.}~\bibnamefont {Maleki}},\
  and\ \bibinfo {author} {\bibfnamefont {R.~G.}\ \bibnamefont {Baraniuk}},\
  }\bibfield  {title} {\bibinfo {title} {{BM3D-PRGAMP}: Compressive phase
  retrieval based on {BM3D} denoising},\ }in\ \href
  {https://ieeexplore.ieee.org/abstract/document/7532810} {\emph {\bibinfo
  {booktitle} {2016 IEEE International Conference on Image Processing
  (ICIP)}}}\ (\bibinfo {organization} {IEEE},\ \bibinfo {year} {2016})\ pp.\
  \bibinfo {pages} {2504--2508}\BibitemShut {NoStop}%
\bibitem [{\citenamefont {Tao}\ \emph {et~al.}(2015)\citenamefont {Tao},
  \citenamefont {Bodington}, \citenamefont {Reinig},\ and\ \citenamefont
  {Kubby}}]{tao2015high}%
  \BibitemOpen
  \bibfield  {author} {\bibinfo {author} {\bibfnamefont {X.}~\bibnamefont
  {Tao}}, \bibinfo {author} {\bibfnamefont {D.}~\bibnamefont {Bodington}},
  \bibinfo {author} {\bibfnamefont {M.}~\bibnamefont {Reinig}},\ and\ \bibinfo
  {author} {\bibfnamefont {J.}~\bibnamefont {Kubby}},\ }\bibfield  {title}
  {\bibinfo {title} {High-speed scanning interferometric focusing by fast
  measurement of binary transmission matrix for channel demixing},\ }\href
  {https://opg.optica.org/oe/fulltext.cfm?uri=oe-23-11-14168&id=318935}
  {\bibfield  {journal} {\bibinfo  {journal} {Opt. Express}\ }\textbf {\bibinfo
  {volume} {23}},\ \bibinfo {pages} {14168} (\bibinfo {year}
  {2015})}\BibitemShut {NoStop}%
\bibitem [{\citenamefont {Goodman}(2007)}]{goodman2007speckle}%
  \BibitemOpen
  \bibfield  {author} {\bibinfo {author} {\bibfnamefont {J.~W.}\ \bibnamefont
  {Goodman}},\ }\href@noop {} {\emph {\bibinfo {title} {Speckle phenomena in
  optics: theory and applications}}}\ (\bibinfo  {publisher} {Roberts and
  Company Publishers},\ \bibinfo {year} {2007})\BibitemShut {NoStop}%
\bibitem [{\citenamefont {Akbulut}\ \emph {et~al.}(2011)\citenamefont
  {Akbulut}, \citenamefont {Huisman}, \citenamefont {van Putten}, \citenamefont
  {Vos},\ and\ \citenamefont {Mosk}}]{akbulut2011focusing}%
  \BibitemOpen
  \bibfield  {author} {\bibinfo {author} {\bibfnamefont {D.}~\bibnamefont
  {Akbulut}}, \bibinfo {author} {\bibfnamefont {T.~J.}\ \bibnamefont
  {Huisman}}, \bibinfo {author} {\bibfnamefont {E.~G.}\ \bibnamefont {van
  Putten}}, \bibinfo {author} {\bibfnamefont {W.~L.}\ \bibnamefont {Vos}},\
  and\ \bibinfo {author} {\bibfnamefont {A.~P.}\ \bibnamefont {Mosk}},\
  }\bibfield  {title} {\bibinfo {title} {Focusing light through random photonic
  media by binary amplitude modulation},\ }\href
  {https://opg.optica.org/oe/fulltext.cfm?uri=oe-19-5-4017&id=210099}
  {\bibfield  {journal} {\bibinfo  {journal} {Opt. Express}\ }\textbf {\bibinfo
  {volume} {19}},\ \bibinfo {pages} {4017} (\bibinfo {year}
  {2011})}\BibitemShut {NoStop}%
\bibitem [{\citenamefont {Yeh}\ \emph {et~al.}(2015)\citenamefont {Yeh},
  \citenamefont {Dong}, \citenamefont {Zhong}, \citenamefont {Tian},
  \citenamefont {Chen}, \citenamefont {Tang}, \citenamefont {Soltanolkotabi},\
  and\ \citenamefont {Waller}}]{yeh2015experimental}%
  \BibitemOpen
  \bibfield  {author} {\bibinfo {author} {\bibfnamefont {L.-H.}\ \bibnamefont
  {Yeh}}, \bibinfo {author} {\bibfnamefont {J.}~\bibnamefont {Dong}}, \bibinfo
  {author} {\bibfnamefont {J.}~\bibnamefont {Zhong}}, \bibinfo {author}
  {\bibfnamefont {L.}~\bibnamefont {Tian}}, \bibinfo {author} {\bibfnamefont
  {M.}~\bibnamefont {Chen}}, \bibinfo {author} {\bibfnamefont {G.}~\bibnamefont
  {Tang}}, \bibinfo {author} {\bibfnamefont {M.}~\bibnamefont
  {Soltanolkotabi}},\ and\ \bibinfo {author} {\bibfnamefont {L.}~\bibnamefont
  {Waller}},\ }\bibfield  {title} {\bibinfo {title} {Experimental robustness of
  fourier ptychography phase retrieval algorithms},\ }\href
  {https://opg.optica.org/oe/fulltext.cfm?uri=oe-23-26-33214&id=333747}
  {\bibfield  {journal} {\bibinfo  {journal} {Opt. Express}\ }\textbf {\bibinfo
  {volume} {23}},\ \bibinfo {pages} {33214} (\bibinfo {year}
  {2015})}\BibitemShut {NoStop}%
\bibitem [{\citenamefont {Trefethen}\ and\ \citenamefont
  {Bau~III}(1997)}]{trefethen1997numerical}%
  \BibitemOpen
  \bibfield  {author} {\bibinfo {author} {\bibfnamefont {L.~N.}\ \bibnamefont
  {Trefethen}}\ and\ \bibinfo {author} {\bibfnamefont {D.}~\bibnamefont
  {Bau~III}},\ }\href@noop {} {\emph {\bibinfo {title} {Numerical linear
  algebra}}},\ Vol.~\bibinfo {volume} {50}\ (\bibinfo  {publisher} {SIAM},\
  \bibinfo {year} {1997})\BibitemShut {NoStop}%
\bibitem [{\citenamefont {Bouchet}\ \emph
  {et~al.}(2021{\natexlab{a}})\citenamefont {Bouchet}, \citenamefont {Rotter},\
  and\ \citenamefont {Mosk}}]{bouchet2021maximum}%
  \BibitemOpen
  \bibfield  {author} {\bibinfo {author} {\bibfnamefont {D.}~\bibnamefont
  {Bouchet}}, \bibinfo {author} {\bibfnamefont {S.}~\bibnamefont {Rotter}},\
  and\ \bibinfo {author} {\bibfnamefont {A.~P.}\ \bibnamefont {Mosk}},\
  }\bibfield  {title} {\bibinfo {title} {Maximum information states for
  coherent scattering measurements},\ }\href
  {https://www.nature.com/articles/s41567-020-01137-4} {\bibfield  {journal}
  {\bibinfo  {journal} {Nat. Phys.}\ }\textbf {\bibinfo {volume} {17}},\
  \bibinfo {pages} {564} (\bibinfo {year} {2021}{\natexlab{a}})}\BibitemShut
  {NoStop}%
\bibitem [{\citenamefont {Bouchet}\ \emph
  {et~al.}(2021{\natexlab{b}})\citenamefont {Bouchet}, \citenamefont
  {Rachbauer}, \citenamefont {Rotter}, \citenamefont {Mosk},\ and\
  \citenamefont {Bossy}}]{bouchet2021optimal}%
  \BibitemOpen
  \bibfield  {author} {\bibinfo {author} {\bibfnamefont {D.}~\bibnamefont
  {Bouchet}}, \bibinfo {author} {\bibfnamefont {L.~M.}\ \bibnamefont
  {Rachbauer}}, \bibinfo {author} {\bibfnamefont {S.}~\bibnamefont {Rotter}},
  \bibinfo {author} {\bibfnamefont {A.~P.}\ \bibnamefont {Mosk}},\ and\
  \bibinfo {author} {\bibfnamefont {E.}~\bibnamefont {Bossy}},\ }\bibfield
  {title} {\bibinfo {title} {Optimal control of coherent light scattering for
  binary decision problems},\ }\href
  {https://journals.aps.org/prl/abstract/10.1103/PhysRevLett.127.253902}
  {\bibfield  {journal} {\bibinfo  {journal} {Phys. Rev. Lett.}\ }\textbf
  {\bibinfo {volume} {127}},\ \bibinfo {pages} {253902} (\bibinfo {year}
  {2021}{\natexlab{b}})}\BibitemShut {NoStop}%
\bibitem [{\citenamefont {Rafayelyan}\ \emph {et~al.}(2020)\citenamefont
  {Rafayelyan}, \citenamefont {Dong}, \citenamefont {Tan}, \citenamefont
  {Krzakala},\ and\ \citenamefont {Gigan}}]{rafayelyan2020large}%
  \BibitemOpen
  \bibfield  {author} {\bibinfo {author} {\bibfnamefont {M.}~\bibnamefont
  {Rafayelyan}}, \bibinfo {author} {\bibfnamefont {J.}~\bibnamefont {Dong}},
  \bibinfo {author} {\bibfnamefont {Y.}~\bibnamefont {Tan}}, \bibinfo {author}
  {\bibfnamefont {F.}~\bibnamefont {Krzakala}},\ and\ \bibinfo {author}
  {\bibfnamefont {S.}~\bibnamefont {Gigan}},\ }\bibfield  {title} {\bibinfo
  {title} {Large-scale optical reservoir computing for spatiotemporal chaotic
  systems prediction},\ }\href
  {https://journals.aps.org/prx/abstract/10.1103/PhysRevX.10.041037} {\bibfield
   {journal} {\bibinfo  {journal} {Phys. Rev. X}\ }\textbf {\bibinfo {volume}
  {10}},\ \bibinfo {pages} {041037} (\bibinfo {year} {2020})}\BibitemShut
  {NoStop}%
\bibitem [{\citenamefont {Engel}\ \emph {et~al.}(2004)\citenamefont {Engel},
  \citenamefont {Mannor},\ and\ \citenamefont {Meir}}]{engel2004kernel}%
  \BibitemOpen
  \bibfield  {author} {\bibinfo {author} {\bibfnamefont {Y.}~\bibnamefont
  {Engel}}, \bibinfo {author} {\bibfnamefont {S.}~\bibnamefont {Mannor}},\ and\
  \bibinfo {author} {\bibfnamefont {R.}~\bibnamefont {Meir}},\ }\bibfield
  {title} {\bibinfo {title} {The kernel recursive least-squares algorithm},\
  }\href {https://ieeexplore.ieee.org/document/1315946} {\bibfield  {journal}
  {\bibinfo  {journal} {IEEE Trans. Signal Process.}\ }\textbf {\bibinfo
  {volume} {52}},\ \bibinfo {pages} {2275} (\bibinfo {year}
  {2004})}\BibitemShut {NoStop}%
\end{thebibliography}%

\end{document}